\documentclass[preprint,12pt]{elsarticle}

\usepackage{bm}
\usepackage{amsmath,amssymb,amsfonts,times}
\usepackage{url}

\usepackage[colorlinks=true, pdfstartview=FitV, linkcolor=red, citecolor=blue, urlcolor=blue]{hyperref}

\usepackage{graphicx}
\usepackage{epsfig}
\usepackage{slashed}
\usepackage{a4wide}
\usepackage{fancyhdr}
\usepackage{subfigure}
\usepackage{color}
\usepackage{rotating}

\usepackage{pdflscape}

\allowdisplaybreaks[1]

\begin{document}

\begin{frontmatter}

\title{Single-particle spectral function of the $\Lambda$ hyperon in finite nuclei}

\author{Isaac Vida\~na}
\address{CFisUC, Department of Physics, University of Coimbra, PT-3004-516
Coimbra, Portugal}

\address{email: ividana@fis.uc.pt}

\begin{abstract}

The spectral function of the $\Lambda$ hyperon in finite nuclei is calculated from the corresponding $\Lambda$ self-energy, which is constructed within a perturbative many-body approach using some of the hyperon-nucleon interactions of the J\"{u}lich and Nijmegen groups. Binding energies, wave functions and disoccupation numbers of different single-particle states are obtained for various hypernuclei from $^5_{\Lambda}$He to $^{209}_{\,\,\,\,\,\Lambda}$Pb. The agreement between the calculated binding energies and experimental data is qualitatively good. The small spin-orbit splitting of the $p-, d-, f-$ and $g-$wave states is confirmed. The discrete and the continuum contributions of the single-$\Lambda$ spectral function are computed. Their appearance is qualitatively similar to that of the nucleons. The $Z$-factor, that measures the importance of correlations, is also calculated. Our results show that its value is relatively large, indicating that the $\Lambda$ hyperon is less correlated than nucleons. This is in agreement with the results obtained by other authors for the correlations of the $\Lambda$ in infinite nuclear matter. The disoccupation numbers are obtained by integrating the spectral function over the energy. 
Our results show that the discrete contribution to the disoccupation number decreases when increasing the momentum of the $\Lambda$. This indicates that, in the production reactions of hypernuclei, the $\Lambda$ hyperon is mostly formed in a quasi-free state.

\end{abstract}
\end{frontmatter}

\date{\today}

\section{Introduction}
\label{sec:intro}

The study of hypernuclei, bound systems composed of nucleons and one or more hyperons, provides a unique tool to extent our present knowledge of conventional nuclear physics to the $SU(3)$-flavour sector \cite{gal16}. 
Hypernuclei were discovered in 1952 with the observation of a hyperfragment in a ballon-flown emulsion stack by Danysz and Pniewski \cite{dapnie52}. These initial cosmic-ray observations of hypernuclei were followed later by pion and proton beam production reactions in emulsions and $^4$He bubble chambers. The weak decay of the $\Lambda$ hyperon into a $\pi^-$ plus a proton was used to identify different $\Lambda$-hypernuclei and to determine the binding energies, spins and lifetimes for hypernuclei up to $A=15$ \cite{juric73, davis91}. Average properties of heavier systems were estimated from spallation experiments, and two double-$\Lambda$ hypernuclei were reported from $\Xi^-$ capture \cite{dl0,dl1,dl1b,dl2,dl3,dl4,dl5}. More systematic investigations of hypernuclei began with the advent of separated $K^-$ beams, which allowed to produce single $\Lambda$-hypernuclei through $(K^-,\pi^-)$ {\it strangeness exchange reactions}, where a neutron hit by a $K^-$ is changed into a $\Lambda$ emitting a $\pi^-$. The analysis of these reactions, initially carried out at CERN and later at BNL, KEK and J-PARC, showed many of the hypernuclear characteristics such as, for instance,  the small spin-orbit strength of the hyperon-nucleon (YN) interaction, or the fact that the $\Lambda$ essentially retains its identity inside the nucleus. The use of $\pi^+$ beams at BNL, KEK and GSI has permitted to perform $(\pi^+,K^+)$ {\it associated production reactions}, where an $s\bar s$ pair is created from the vacuum, and a $K^+$ and a $\Lambda$ are produced in the final state. The {\it electro-production} of hypernuclei at JLAB and MAMI-C by means of the reaction $(e,e'K^+)$ provides a high precision tool for the study of hypernuclear spectroscopy \cite{hugenford94} due to the excellent spatial and energy resolution of the electron beams. The HypHI collaboration at FAIR/GSI has recently proposed a completely new and alternative way to produce hypernuclei by using stable and unstable heavy ion beams \cite{hypHI}. A first experiment has been already performed using a $^6$Li beam on a $^{12}$C target at 2 A GeV, in which the $\Lambda$ and the $^3_{\Lambda}$H and $^4_{\Lambda}$H hypernuclei have been observed \cite{rappold13}. 

Nowadays more than 40 single $\Lambda$-hypernuclei, and few double-$\Lambda$ \cite{dl0,dl1,dl1b,dl2,dl3,dl4,dl5,nagara} and single-$\Xi$ \cite{khaustov00,nakazawa15} ones have been identified. On the contrary, it has not been possible to prove without any ambiguity the existence of $\Sigma$-hypernuclei (see {\it e.g.,} Refs.\  \cite{bertini80,bertini84,bertini85,piekarz82,yamazaki85,tang88,bart99,hayano89,nagae98}) which suggest that the $\Sigma$-nucleon interaction is most probably repulsive \cite{dgm89,batty1,batty2,batty3,mares95,dabrowski99,noumi02,saha04,harada05,harada06}. 

A simple theoretical description of $\Lambda$-hypernuclei consist of an ordinary nuclei with the $\Lambda$ sitting in the single-particle states of an effective $\Lambda$-nucleus mean field potential. Several approaches, based on this simple description, have been used to study the properties of the $\Lambda$ in finite nuclei.  Woods-Saxon potentials, for instance, have been traditionally used to describe, in a shell model picture, the single-particle properties of the $\Lambda$ from medium to heavy hypernuclei \cite{ws1,ws1b,ws2,ws3}. Density dependent effects and non-localities have been included in non-relativistic Hartree--Fock calculations with Skyrme type YN interactions in order to improve the overall fit of the single-particle energies 
\cite{skhf1,skhf2,skhf3,skhf4,skhf4b,skhf5,skhf5b,skhf5c,skhf5d}. Hypernuclear structure calculations have been also performed on the framework of relativistic mean field theory 
\cite{rmf1,rmf2,rmf3,rmf4,rmf5,rmf6,rmf7,rmf8,rmf8b,rmf9} and Dirac phenomenology \cite{dirac1,dirac2}. Several hypernuclear structure studies based on {\it ab-initio} approaches exist in the literature
\cite{g1,g1b,g1c,g1d,g2,g3,morten96,vidana98,vidana00}. In these studies the single-particle properties of the $\Lambda$ in the hypernucleus are derived from effective YN $G$-matrices built from bare YN interactions which describe the scattering data in free space. Recently, a Quantum Monte Carlo calculation of single- and double-$\Lambda$ hypernuclei has also been done using two- and three-body forces between the $\Lambda$ and the nucleons \cite{lonardoni1,qmchyp}.

In most of these approaches, the quality of the description of hypernuclei relies on the validity of the mean field picture. However, the correlations induced by the YN interaction  
can substantially change this picture and, therefore, should not be ignored. Whereas the correlations of nucleons in nuclear matter and finite nuclei have been extensively studied by many authors (see {\it e.g.,} Refs.\ \cite{omar89,omar90,omar92,dickhoff92,omar93,mudick94,vanneck94,artur95,ciofi95,herbert96,artur00,barbieri04,tobias04,tobias05,omar08,rios09,rios09b,artur11,carbone13} and references therein), the correlations of hyperons have not received so much attention up to date. To the best of our knowledge, the effect of the $\Lambda$ correlations in nuclear matter, beyond the mean field description, have been only studied by Robertson and Dickhoff \cite{robertson04,robertson04b}. These authors used the Green's function formalism \cite{fetter,dickhoffbook} to study for the fist time the propagation of a $\Lambda$ in nuclear matter. They calculated the spectral function and quasi-particle parameters of the $\Lambda$ finding results qualitatively similar to those of the nucleons, and showed that the  $\Lambda$ is, in general, less correlated than the nucleons. 

The knowledge of the single-particle spectral function of the $\Lambda$ in finite nuclei is fundamental not only to determine up to which extent the mean field description of hypernuclei is valid, but also for a proper description of the cross section of the different production mechanism of hypernuclei. Information on the single-$\Lambda$ spectral function can be obtained from a combined analysis of data provided by {\it e.g.,} $(e,e'K^+)$ reactions or other experiments with theoretical calculations. However, as far as we know, until now the single-$\Lambda$ spectral function in finite nuclei has never been calculated. The scope of this work is to determine it for a variety of hypernuclei. To such end, we use a perturbative many-body approach to determine first the $\Lambda$ self-energy in finite nuclei from which we can then obtain the single-$\Lambda$ spectral function and the $\Lambda$ single-particle bound states for the different hypernuclei.  

The manuscript is organized in the following way. In Sec.\ \ref{sec:sf} we describe in detail the method employed to determine the $\Lambda$ spectral function in finite nuclei. Results for the $\Lambda$ self-energy, single-particle bound states, spectral function and disoccupation numbers in several hypernuclei from $^5_{\Lambda}$He to $^{209}_{\,\,\,\,\,\Lambda}$Pb are presented and discussed in Sec.\ \ref{sec:res}. Finally, our main conclusions are summarized in Sec. \ref{sec:sumcon}.


\section{Evaluation of the single-$\Lambda$ spectral function in finite nuclei}
\label{sec:sf}

In this section we describe the calculation of  the single-$\Lambda$ spectral function in a hypernucleus. The starting 
point of this calculation is a nuclear matter YN $G$-matrix evaluated in momentum space at fixed nuclear matter density, center-of-mass momentum and starting energy. This nuclear matter $G$-matrix is then used to construct a finite nucleus YN $G$-matrix from which we can obtain the self-energy of the $\Lambda$ in the Brueckner--Hartree--Fock (BHF) approximation, and the corresponding spectral function, for different single-particle bound and scattering states of an effective hyperon-nucleus potential in several hypernuclei. The calculation is done using some of the YN interactions of the J\"{u}lich \cite{juelich,juelich2} and the Nijmegen \cite{nijmegen,nijmegen2,nijmegen3} groups. The description of our calculation is presented in the following after some 
few general remarks on the single-particle propagator and the spectral function.

\subsection{General remarks}

It is well known in quantum many-body theory that the propagation of a particle or a hole with incoming (outgoing) quantum numbers $\alpha \, (\beta)$ and energy $\omega$ 
that is added to a given $N-$particle system is described by the
single-particle Green's function (or propagator) $g_{\alpha\beta} (\omega)$, which can be obtained by solving the Dyson equation \cite{fetter,dickhoffbook}
\begin{equation}
g_{\alpha\beta}(\omega)=g^{(0)}_{\alpha\beta}(\omega) + \sum_{\gamma}\sum_{\eta}g^{(0)}_{\alpha\gamma}(\omega)\Sigma_{\gamma\eta}(\omega)g_{\eta\beta}(\omega) \ ,
\label{eq:de}
\end{equation}
where 
\begin{equation}
g^{(0)}_{\alpha\beta}(\omega)
=\langle\Phi^{N}_0|\hat c_\alpha \frac{1}{\omega-(\hat H_0-E_{\Phi^N_{0}})+i\eta}\hat c^\dag_\beta |\Phi^{N}_0\rangle
\mp \langle\Phi^{N}_0|\hat c^\dag_\beta \frac{1}{\omega-(E_{\Phi^N_{0}}-\hat H_0)-i\eta}\hat c_\alpha |\Phi^{N}_0\rangle 
\label{eq:fspp}
\end{equation}
is the free single-particle propagator with $\hat H_0$ being the Hamiltonian of the non-interacting $N$-particle system, $|\Phi^{N}_0\rangle$ its corresponding 
ground state of energy $E_{\Phi^N_{0}}$ and the minus (plus) sign in front of  the second term is for bosons (fermions). $\Sigma_{\alpha\beta}(\omega)$ is the proper or irreducible self-energy that from now on will be simply referred as self-energy. 

Particularly useful is the so-called Lehmann representation of the propagator
\begin{equation}
g_{\alpha\beta}(\omega)
=\sum_{m}\frac{\langle\Psi^{N}_0|\hat c_\alpha|\Psi^{N+1}_m\rangle \langle\Psi^{N+1}_m|\hat c^\dag_\beta|\Psi^{N}_0\rangle }{\omega-(E^{N+1}_m-E^{N}_0)+i\eta}
\mp \sum_{n}\frac{\langle\Psi^{N}_0|\hat c^\dag_\beta|\Psi^{N-1}_n\rangle \langle\Psi^{N-1}_n|\hat c_\alpha|\Psi^{N}_0\rangle }{\omega-(E^{N}_0-E^{N-1}_n)-i\eta} \ ,
\label{eq:lr}
\end{equation}
where $|\Psi^{N}_0\rangle$ is the ground state of the $N-$particle interacting system and $|\Psi^{N\pm 1}_k\rangle$ are the eigenstates of the Hamiltonian of 
the $(N\pm 1)-$particle one. Alternatively, the Lehmann representation of the propagator can be written
in the form
\begin{equation}
g_{\alpha\beta}(\omega)=\int_{E^{N+1}_0-E^{N}_0}^{\infty} d\omega' \frac{S^p_{\alpha\beta}(\omega')}{\omega-\omega'+i\eta}
+\int_{-\infty}^{E^{N}_0-E^{N-1}_0} d\omega' \frac{S^h_{\alpha\beta}(\omega')}{\omega-\omega'-i\eta} \ ,
\label{eq:lr2}
\end{equation}
where $S^p_{\alpha\beta}(\omega)$ and $S^h_{\alpha\beta}(\omega)$ are, respectively, the particle and the hole parts of the single-particle spectral function which are defined as
\begin{equation}
S^p_{\alpha\beta}(\omega)=\sum_{m} \langle\Psi^{N}_0|\hat c_\alpha|\Psi^{N+1}_m\rangle \langle\Psi^{N+1}_m|\hat c^\dag_\beta|\Psi^{N}_0\rangle \delta(\omega-(E^{N+1}_m-E^{N}_0)), 
\,\,\,\,\, \omega > E^{N+1}_0-E^{N}_0
\label{eq:sp}
\end{equation}
and
\begin{equation}
S^h_{\alpha\beta}(\omega)=\mp \sum_{n} \langle\Psi^{N}_0|\hat c^\dag_\beta|\Psi^{N-1}_n\rangle \langle\Psi^{N-1}_n|\hat c_\alpha|\Psi^{N}_0\rangle \delta(\omega-(E^{N}_0-E^{N-1}_n)),
\,\,\,\,\, \omega < E^{N}_0-E^{N-1}_0 \ .
\label{eq:sh}
\end{equation}
We note that $S^p_{\alpha\alpha}(\omega)$ and $S^h_{\alpha\alpha}(\omega)$ give, respectively, the probability density of adding or removing 
a particle with quantum numbers $\alpha$ to the ground state of the $N-$particle system, and finding the resulting $(N+1)-$ or $(N-1)-$particle one in an excited state 
of energy $\omega-(E^{N+1}_0-E^{N}_0)$ (with $\omega > E^{N+1}_0-E^{N}_0$) or $(E^{N}_0-E^{N-1}_0) -\omega$ (with $\omega < E^{N}_0-E^{N-1}_0$). 

It is easy to see from Eqs.\ (\ref{eq:sp}) and (\ref{eq:sh}) that $S^p_{\alpha\beta}(\omega)$ and $S^h_{\alpha\beta}(\omega)$ fulfill the following sum rule
\begin{equation}
\int_{E^{N+1}_0-E^{N}_0}^{\infty} d\omega S^p_{\alpha\beta}(\omega)
+\int_{-\infty}^{E^{N}_0-E^{N-1}_0} d\omega S^h_{\alpha\beta}(\omega)
=\langle \Psi^{N}_0| [\hat c_\alpha, \hat c^\dag_\beta]_\mp|\Psi^{N}_0 \rangle=\delta_{\alpha\beta} \ ,
\label{eq:sumrule0}
\end{equation}
where, in the last equality, it has been used $[\hat c_\alpha, \hat c^\dag_\beta]_\mp=\delta_{\alpha\beta}$ and assumed $\langle \Psi^{N}_0|\Psi^{N}_0 \rangle=1$.

After these general remarks let us consider the particular case of a $\Lambda$ hyperon that is added to a pure nucleonic system such as {\it e.g.,} infinite nuclear matter or an ordinary nuclei. It is clear that, since there are no other $\Lambda$'s in the $N-$particle pure nucleonic system, the $\Lambda$ can only be added to it and, therefore, the hole part of its spectral function is zero. The Lehmann representation of the single-$\Lambda$ propagator is then simply given by
\begin{equation}
g^{\Lambda}_{\alpha\beta}(\omega)=\int_{E^{N+\Lambda}_0-E^{N}_0}^{\infty} d\omega' \frac{S^{\Lambda p}_{\alpha\beta}(\omega')}{\omega-\omega'+i\eta} \ .
\label{eq:lrl}
\end{equation}
Similarly, the sum rule expressed by Eq.\ (\ref{eq:sumrule0}) is simplified, reading in this case
\begin{equation}
\int_{E^{N+\Lambda}_0-E^{N}_0}^{\infty} d\omega S^{\Lambda p}_{\alpha\beta}(\omega)=
\langle \Psi^{N}_0| \hat c^\Lambda_\alpha\hat c^{\Lambda\dag}_\beta|\Psi^{N}_0 \rangle =
\langle \Psi^{N}_0| \delta_{\alpha\beta}-\hat c^{\Lambda\dag}_\beta\hat c^\Lambda_\alpha|\Psi^{N}_0 \rangle =
\delta_{\alpha\beta} \ ,
\label{eq:sumrule1}
\end{equation}
where we have used that the action of the operator $\hat c^\Lambda_\alpha$, which annihilates a $\Lambda$ hyperon with quantum numbers $\alpha$, on the pure nucleonic state $|\Psi^{N}_0 \rangle$ is zero because this state does not contains any $\Lambda$.

\subsection{$\Lambda$ self-energy in finite nuclei}

Our calculation of the $\Lambda$ self-energy is based on a method that was originally developed to study the properties of the nucleon \cite{borromeo92} and the $\Delta$ isobar \cite{morten94} in finite nuclei, and was later extended to study those of the $\Lambda$ and $\Sigma$ hyperons \cite{morten96,vidana98,vidana00}. In the following we present a general description of this method and refer the interested reader to these works, in  particular to Refs.\  \cite{morten96,vidana98,vidana00},  for specific details of the calculation.

The evaluation of the $\Lambda$ self-energy in finite nucleus starts with the construction of all the YN $G$-matrices which describe the in-medium interaction between a hyperon (Y$=\Lambda,\Sigma$) and a nucleon in infinite nuclear matter. The $G$-matrices 
are obtained by solving the coupled-channel Bethe--Goldstone equation which schematically reads
\begin{eqnarray}
\langle Y'N'|G| YN\rangle &=& \langle Y'N' | V | YN \rangle \,\,
+\sum_{Y''N''=\Lambda N, \Sigma N} \langle Y'N' |V|Y''N''\rangle \nonumber \\
&&\times\frac{Q_{Y''N''}}{\Omega-\varepsilon_{Y''}-\varepsilon_{N''}+i\eta} 
\langle Y''N''|G|YN\rangle \ ,
\label{eq:gfn}
\end{eqnarray}
where $V$ is the bare YN interaction, $Q_{Y"N"}$ is the Pauli operator, that prevents the nucleon in the intermediate
state Y$''$N$''$ to be scattered below its Fermi momentum $k_{F_N}$, and 
$\Omega$ is the so-called starting energy which is 
the sum of the non-relativistic single-particle energies of the interacting hyperon and nucleon. We note here that the 
Bethe--Goldstone equation has been solved in momentum space in the partial wave basis
$|YN\rangle \equiv |K{\cal L}qLSJ{\cal J}TM_T\rangle$ where $K (q)$ and ${\cal L} (L)$ are respectively the center-of-mass (relative) momentum and orbital angular momentum, $S$ is the total spin, ${\cal J}$ is the total angular momentum,
$T$ and $M_T$ are the total isospin and its third component, and $\vec J=\vec L+\vec S$. We note also that 
when solving it, the so-called discontinuous prescription has been adopted, {\it i.e.,} the single-particle energy of the hyperon ($\varepsilon_{Y''}$) and the nucleon ($\varepsilon_{N''}$) in the intermediate state Y$''$N$''$ is taken simply as the sum of the non-relativistic kinetic energy plus the mass of the corresponding baryon. Finally, we should mention that
our calculation has been done at nuclear matter density $\rho=0.17$ fm$^{-3}$, zero center-of-mass momentum and $\Omega=m_N+m_\Lambda-80$ MeV, where $-80$ MeV is an averaged value for the sum of the single-particle mean fields of the nucleon ($U_N(k=k_{F_N}) \approx -50$ MeV) and the $\Lambda$ ($U_\Lambda(k=0) \approx -30$ MeV) at this density. The dependence of our results on the values of the nuclear matter density $\rho$ and the starting energy $\Omega$ is weak as it is shown {\it e.g.,} in Refs.\ \cite{vidana98,vidana00}.

The finite nucleus YN $G$-matrix, $G_{FN}$, can be obtained, in principle, by solving the Bethe--Goldstone equation directly in the finite nucleus (see {\it e.g.,} Refs.\ \cite{g2,g3}) which is formally identical to Eq.\ (\ref{eq:gfn}), being the only differences the Pauli operator and the energy denominator which in this case are the ones corresponding to the finite nucleus case. Alternatively, one can find $G_{FN}$ by relating it to the nuclear matter $G$-matrix already obtained. Eliminating the bare interaction $V$ in both finite nucleus and nuclear matter Bethe--Goldstone equations, $G_{FN}$ can be written in terms of $G$ through the following integral equation, written in a simplyfied form as
\begin{eqnarray}
G_{FN}&=&G+G\left[\frac{Q_{FN}}{E_{FN}}-\frac{Q}{E}\right]G_{FN} \nonumber \\
&=&G+G\left[\frac{Q_{FN}}{E_{FN}}-\frac{Q}{E}\right]G 
+G\left[\frac{Q_{FN}}{E_{FN}}-\frac{Q}{E}\right]G\left[\frac{Q_{FN}}{E_{FN}}-\frac{Q}{E}\right]G 
+ \cdot \cdot \cdot \ ,
\label{eq:gfng}
\end{eqnarray}
where the difference $Q_{FN}/E_{FN}-Q/E$ (being $Q_{FN}$, $E_{FN}$, $Q$ and $E$ the corresponding finite nucleus and nuclear matter Pauli operators and energy denominators) accounts for the relevant intermediate particle-particle states.
This difference has been shown to be quite small \cite{morten96,vidana98,vidana00,borromeo92,morten94}  and, therefore, in all practical calculations $G_{FN}$ can be well approximated by truncating the expansion (\ref{eq:gfng}) up second order in 
the nuclear matter $G$-matrix. Therefore, we have
\begin{equation}
G_{FN} \simeq G+G\left[\frac{Q_{FN}}{E_{FN}}-\frac{Q}{E}\right]G \ .
\label{eq:g2nd}
\end{equation}

\begin{figure}[t]
\begin{center}
\includegraphics[width=12.5cm]{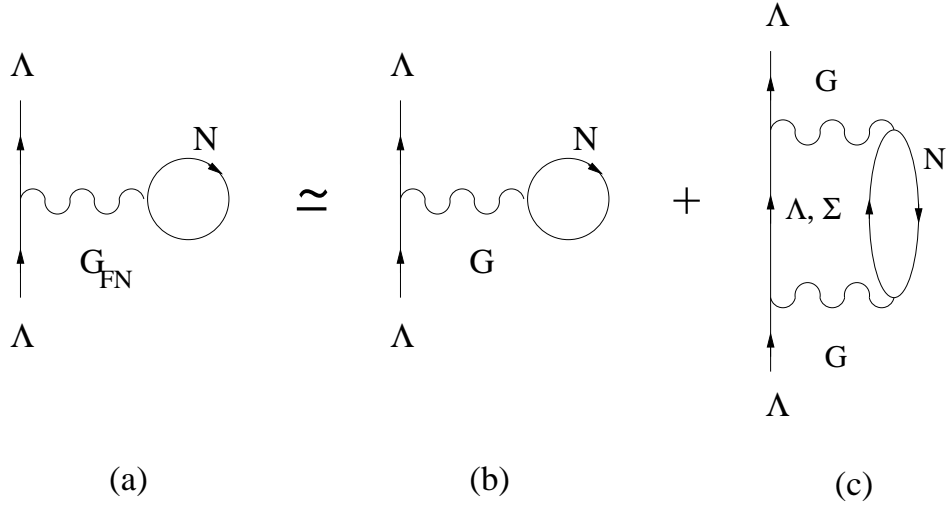}
\caption{Brueckner--Hartree--Fock approximation of the finite nucleus $\Lambda$ self-energy (diagram (a)), split into the sum of a
first-order contribution (diagram (b)) and a second-order 2p1h correction (diagram (c)).}
\label{fig:selfener}
\end{center}
\end{figure}

Using now $G_{FN}$ as an effective YN interaction, we can calculate the finite nucleus $\Lambda$ self-energy in the BHF approximation (see diagram (a) of Fig.\ \ref{fig:selfener}). According to Eq.\ (\ref{eq:g2nd}) this approximation can be split into the sum of two contributions: the one of diagram (b), which represents the first-order term on the right-hand side of Eq.\ (\ref{eq:g2nd}), and that of diagram (c), which stands for the so-called {\it two-particle-one-hole} ($2p1h$) correction. In the following we show the explicit expressions of both contributions without going into many details of their derivation which can be found in Refs.\ \cite{morten96,vidana98,vidana00}. 

Diagram (b) of Fig.\ \ref{fig:selfener} gives the following real and energy-independent contribution to the $\Lambda$ self-energy
\begin{eqnarray}
{\cal V}_{1}(k_\Lambda,k'_\Lambda,l_\Lambda, j_\Lambda)&=&\frac{1}{2j_\Lambda+1}\sum_{\cal J}\sum_{n_hl_hj_ht_{z_h}}
(2{\cal J}+1) \nonumber \\
&&\times \langle (k'_\Lambda l_\Lambda j_\Lambda)(n_hl_hj_ht_{z_h}){\cal J}| G |
(k_\Lambda l_\Lambda j_\Lambda)(n_hl_hj_ht_{z_h}){\cal J} \rangle \ ,
\label{eq:diagb}
\end{eqnarray}
where the incoming (outgoing) $\Lambda$ and the nucleon hole are respectively taken as plane wave and harmonic oscillator states with quantum numbers $k_\Lambda (k'_\Lambda)l_\Lambda j_\Lambda$ and $n_hl_hj_ht_{z_h}$.  The total angular momentum of the $\Lambda$ nucleon hole pair is ${\cal \vec J}=\vec j_\Lambda+\vec j_h$ in the laboratory frame.  Note that in the above expression and in the following ones neither the z-component of the $\Lambda$ isospin $t_{z_\Lambda}=0$ nor its total spin and that of the nucleons is shown explicitly.

The contribution of diagram (c) is the sum of two terms. The first of them is the  term $G\left(Q_{FN}/E_{FN}\right)G$ in Eq.\ (\ref{eq:g2nd}) which gives rise to an imaginary part in the $\Lambda$ self-energy which depends explicitly on the energy of the hyperon and reads
\begin{eqnarray}
{\cal W}_{2p1h}(k_\Lambda,k'_\Lambda, l_\Lambda, j_\Lambda,\omega)&=&-\frac{\pi}{2j_\Lambda+1}\sum_{n_hl_hj_ht_{z_h}}
\sum_{{\cal L}LSJ{\cal J}}\sum_{Y'=\Lambda\Sigma}\int dq q^2\int dK K^2 (2{\cal J}+1) \nonumber \\
&&\times
\langle (k'_\Lambda l_\Lambda j_\Lambda)(n_hl_hj_ht_{z_h}){\cal J}|G|
K{\cal L}qLSJ{\cal J}TM_T\rangle \nonumber \\
&&\times
\langle K{\cal L}qLSJ{\cal J}TM_T|G|
(k_\Lambda l_\Lambda j_\Lambda )(n_hl_hj_ht_{z_h}){\cal J}\rangle \nonumber \\
&&\times
\delta\left(\omega+\varepsilon_h-\frac{\hbar^2K^2}{2(m_N+m_{Y'})}-\frac{\hbar^2q^2(m_N+m_{Y'})}{2m_Nm_{Y'}}-m_{Y'}+m_{\Lambda}\right) \ , \nonumber \\
\label{eq:diagc1}
\end{eqnarray}
where $\omega$ is the energy of the $\Lambda$ measured with respect to its rest mass. The quantum numbers 
$K,{\cal L},q,L,S,J,{\cal J},T$ and $M_T$ have been defined before, and the energies of the nucleon hole states $\varepsilon_h$ have been taken equal to the experimental single-particle ones of the nuclei studied. One can see from the delta function that ${\cal W}_{2p1h}$ is different from zero only for positive values of $\omega$. The contribution of this term to the real part of the $\Lambda$ self-energy can be obtained through the following dispersion relation
\begin{equation}
{\cal V}^{(1)}_{2p1h}(k_\Lambda,k'_\Lambda,l_\Lambda, j_\Lambda,\omega) =\frac{1}{\pi}{\cal P}\int_{-\infty}^{\infty}d\omega' 
\frac{{\cal W}_{2p1h}(k_\Lambda,k'_\Lambda,l_\Lambda, j_\Lambda,\omega')}{\omega'-\omega} \ ,
\label{eq:diagc2}
\end{equation}
where ${\cal P}$ stands for a principal value integral.

The second contribution of diagram (c) is that of the term $G\left(Q/E\right)G$ in Eq.\ (\ref{eq:g2nd}). This term
gives also a real and energy-independent contribution to the $\Lambda$ self-energy, and avoids double counting over intermediate Y$'$N states that are already contained in the nuclear matter $G$-matrix of the contribution of diagram (b). It reads
\begin{eqnarray}
{\cal V}^{(2)}_{2p1h}(k_\Lambda,k'_\Lambda,l_\Lambda, j_\Lambda)&=&\frac{1}{2j_\Lambda+1}\sum_{n_hl_hj_ht_{z_h}}
\sum_{{\cal L}LSJ{\cal J}}\sum_{Y'=\Lambda\Sigma}\int dq q^2\int dK K^2 (2{\cal J}+1) \nonumber \\
&&\times
\langle (k'_\Lambda l_\Lambda j_\Lambda)(n_hl_hj_ht_{z_h}){\cal J}|G|
K{\cal L}qLSJ{\cal J}TM_T\rangle \nonumber \\
&&\times
\langle K{\cal L}qLSJ{\cal J}TM_T|G|
(k_\Lambda l_\Lambda j_\Lambda)(n_hl_hj_ht_{z_h}){\cal J}\rangle \nonumber \\
&&\times
Q_{Y'N}\left(\Omega-\frac{\hbar^2K^2}{2(m_N+m_{Y'})}-\frac{\hbar^2q^2(m_N+m_{Y'})}{2m_Nm_{Y'}}
-m_{Y'}+m_{\Lambda}\right)^{-1} \ , \nonumber \\
\label{eq:diagc3}
\end{eqnarray}
where $Q_{Y'N}$ is the Pauli operator and $\Omega$ is the starting energy introduced before in Eq.\ (\ref{eq:gfn}).

Further details on the evaluation of the $\Lambda$ self-energy, such as the transformation from the basis $|(k_\Lambda l_\Lambda j_\Lambda)(n_hl_hj_ht_{z_h}){\cal J}\rangle$ to the partial wave one $|K{\cal L}qLSJ{\cal J}TM_T\rangle$ used to solve Eq.\ (\ref{eq:gfn}), or the orthogonalization procedure  of the occupied nucleon states with the distorted plane wave associated with the nucleon in the intermediate state of diagram (c), can be found in Refs.\ \cite{morten96,vidana98,vidana00,borromeo92,morten94}.

Summarizing, the BHF approximation of the self-energy of a $\Lambda$ with incoming (outgoing) momentum $k_\Lambda (k'_\Lambda)$, orbital angular momentum $l_\Lambda$, and total angular momentum $j_\Lambda$
is given by
\begin{equation}
\Sigma_{l_\Lambda j_\Lambda}(k_\Lambda,k'_\Lambda,\omega)={\cal V}_{l_\Lambda j_\Lambda}(k_\Lambda,k'_\Lambda, \omega)+i{\cal W}_{l_\Lambda j_\Lambda}(k_\Lambda,k'_\Lambda,\omega) \ ,
\label{eq:lse}
\end{equation}
where the real and imaginary parts are
\begin{equation}
{\cal V}_{l_\Lambda j_\Lambda}(k_\Lambda,k'_\Lambda, \omega)=
{\cal V}_{1}(k_\Lambda,k'_\Lambda,l_\Lambda, j_\Lambda)
+
{\cal V}^{(1)}_{2p1h}(k_\Lambda,k'_\Lambda,l_\Lambda, j_\Lambda,\omega)
-
{\cal V}^{(2)}_{2p1h}(k_\Lambda,k'_\Lambda,l_\Lambda, j_\Lambda)
\label{eq:lser}
\end{equation}
and 
\begin{equation}
{\cal W}_{l_\Lambda j_\Lambda}(k_\Lambda,k'_\Lambda,\omega)=
{\cal W}_{2p1h}(k_\Lambda,k'_\Lambda,l_\Lambda, j_\Lambda,\omega) \ .
\label{eq:lsei}
\end{equation}

\subsection{$\Lambda$ particle spectral function}

In any production mechanism of single-$\Lambda$ hypernuclei a $\Lambda$ can be formed either in a bound ($\omega<0$) or in a scattering ($\omega>0$) state. Therefore, its particle spectral function is the sum of a discrete
and a continuum contribution, whose diagonal parts give, respectively, the probability density of adding the $\Lambda$ in a bound or a scattering state of the corresponding hyperon-nucleus potential.

The discrete contribution of the $\Lambda$ particle spectral function is a weighted delta function located at the energy corresponding to the $\Lambda$ bound state. To calculate it, we need first to determine this state. This can be done by using the real part of the $\Lambda$ self-energy as an effective single-particle hyperon-nucleus potential in the Schr\"{o}dinger equation. Following the procedure outlined in Refs.\ 
\cite{morten96,vidana98,vidana00,borromeo92, morten94}, we solve it by diagonalizing the corresponding single-particle Hamiltonian in a complete and orthonormal set of regular basis functions within a spherical box of radius $R_{box}$ given in coordinate representation by 
\begin{equation}
\Phi_{n l_\Lambda j_\Lambda m_{j_\Lambda}}(\vec r)=\langle\vec r| k_n l_\Lambda j_\Lambda m_{j_\Lambda}\rangle = N_{n l_\Lambda}\,j_{l_\Lambda}(k_nr)\psi_{l_\Lambda j_\Lambda m_{j_\Lambda}}(\theta,\phi) \ ,
\label{eq:basis}
\end{equation}
where $N_{n l_\Lambda}$ is a normalization constant
\begin{eqnarray}
N_{n l_\Lambda} = \left\{
\begin{array}{c}
\frac{\sqrt{2}}{\sqrt{R_{box}^3}j_{l_\Lambda-1}(k_nR_{box}) } \,\,\,\,\, \mbox{for} \,\,\,\,\, l_\Lambda > 0 \\
\frac{n\pi\sqrt{2}}{\sqrt{R_{box}^3}} \,\,\,\,\,\,\,\,\,\,\,\,\,\,\, \,\,\,\,\,\,\,\,\,\,\,\,\,\,\,\,\,\,\,\,\mbox{for} \,\,\,\,\, l_\Lambda = 0 \ ,
\end{array}
\right. 
\end{eqnarray}
$\psi_{l_\Lambda j_\Lambda m_{j_\Lambda}}(\theta,\phi)$ represents the spherical harmonics including the spin degree of freedom, and $j_{l_\Lambda}(k_nr)$ denote the spherical Bessel functions for the discrete momenta $k_n$ which can be obtained from the condition 
\begin{equation}
j_{l_\Lambda}(k_nR_{box})=0 \ . 
\label{eq:bessel}
\end{equation}

To guarantee the independence of the results on $R_{box}$, its value should be larger than the radius of the nucleus considered. Typically $R_{box}$ is chosen around 20 fm or larger. The resulting eigenvalue problem reads
\begin{equation}
\sum_{i=1}^{N_{max}} \left[ \frac{\hbar^2k_i^2}{2m_{\Lambda}}
+{\cal V}_{l_\Lambda j_\Lambda}(k_n,k_i, \omega=\varepsilon_{l_\Lambda j_\Lambda})\right]\Psi_{i l_\Lambda j_\Lambda m_{j_\Lambda}}= \varepsilon_{l_\Lambda j_\Lambda}\Psi_{n l_\Lambda j_\Lambda m_{j_\Lambda}}\ ,
\label{eq:eigenvalue}
\end{equation}
where the maximum number of basis states in the box, $N_{max}$, is typically restricted to 20 or 30, and $\Psi_{n l_\Lambda j_\Lambda m_{j_\Lambda}}\equiv\langle k_n l_\Lambda j_\Lambda m_{j_\Lambda}| \Psi \rangle$ denotes the projection of the state $| \Psi \rangle$ on the basis $| k_n l_\Lambda j_\Lambda m_\Lambda \rangle$.
Note that a self-consistent procedure is required for each eigenvalue, {\it i.e.,} the $\Lambda$ self-energy should be evaluated at each step of the iterative process at the energy of the resulting eigenvalue until convergence is achieved. 

Once the solution of the Schr\"{o}dinger equation is found we can finally obtain the contribution of this bound state 
to the diagonal part of the $\Lambda$ particle spectral function for the set  of discrete momenta $k_\Lambda=k_n$ simply as 
\begin{equation}
S^{p(d)}_{l_\Lambda j_\Lambda }(k_n, \omega) = Z_{l_\Lambda j_\Lambda}
|\langle k_n l_\Lambda j_\Lambda m_{j_\Lambda}| \Psi \rangle|^2
\delta(\omega-\varepsilon_{l_\Lambda j_\Lambda})\ ,
\label{eq:sfld}
\end{equation}
where 
\begin{equation}
Z_{l_\Lambda j_\Lambda}=\left(1-\frac{\partial \langle \Psi |
\Sigma_{l_\Lambda j_\Lambda}(\omega)| \Psi \rangle}{\partial \omega}\Big|_{\omega=\varepsilon_{l_\Lambda j_\Lambda}} \right)^{-1}
\label{eq:specf}
\end{equation}
is the so-called $Z$-factor \cite{dickhoff92} with the expectation value of the self-energy given by
\begin{equation}
\langle \Psi | \Sigma_{l_\Lambda j_\Lambda}(\omega)|\Psi \rangle = \sum_{i,n=1}^{N_{max}} \Psi^*_{i l_\Lambda j_\Lambda m_{j_\Lambda}}
\Sigma_{l_\Lambda j_\Lambda}(k_i,k_n,\omega) \Psi_{n l_\Lambda j_\Lambda m_{j_\Lambda}} \ .
\label{eq:ev}
\end{equation}
The contribution from the bound state $\varepsilon_{l_\Lambda j_\Lambda}$ to the total spectral strength is obtained by summing Eq.\ (\ref{eq:sfld}) over all discrete momenta $k_n$. Since 
$\sum_{n=1}^{N_{max}}|\langle k_n l_\Lambda j_\Lambda m_{j_\Lambda} | \Psi \rangle|^2=1$,
this contribution simple reads
\begin{equation}
S^{p(d)}_{l_\Lambda j_\Lambda}(\omega) =Z_{l_\Lambda j_\Lambda}
\delta(\omega-\varepsilon_{l_\Lambda j_\Lambda})\ .
\label{eq:ssfld2}
\end{equation}

To determine the continuum contribution of the $\Lambda$ particle spectral function, first we should obtained the complete reducible $\Lambda$ self-energy. This can be done by iterating the irreducible one to all orders
\begin{equation}
\Sigma^{red}_{l_\Lambda j_\Lambda}(k_\Lambda,k'_\Lambda, \omega)
=
\Sigma_{l_\Lambda j_\Lambda}(k_\Lambda,k'_\Lambda, \omega)
+\int dq_\Lambda q^2_\Lambda \Sigma_{l_\Lambda j_\Lambda}(k_\Lambda,q_\Lambda, \omega) 
g_\Lambda^{(0)}(q_\Lambda,\omega)\Sigma^{red}_{l_\Lambda j_\Lambda}(q_\Lambda,k'_\Lambda, \omega) \ ,
\label{eq:deq2}
\end{equation}
where $g_\Lambda^{(0)}(q_\Lambda,\omega)=\left(\omega-\hbar^2q^2_\Lambda/2m_\Lambda+i\eta\right)^{-1}$. Then, the single-$\Lambda$ propagator can be derived from the following form of the Dyson equation \cite{dickhoffbook,mahaux91,dussan14}
\begin{equation}
g^{\Lambda}_{l_\Lambda j_\Lambda }(k_\Lambda,k'_\Lambda, \omega)
=
\frac{\delta(k_\Lambda-k'_\Lambda)}{k^2}g_\Lambda^{(0)}(k_\Lambda,\omega) 
+g_\Lambda^{(0)}(k_\Lambda,\omega)\Sigma^{red}_{l_\Lambda j_\Lambda}(k_\Lambda,k'_\Lambda, \omega)g_\Lambda^{(0)}(k'_\Lambda,\omega) \ .
\label{eq:slp2}
\end{equation}
Once the single-$\Lambda$ propagator is known, the continuum contribution of the $\Lambda$ particle spectral function can be finally obtained from the discontinuity of the particle part of the propagator across the branch cut that runs below the real axis for positive energies \cite{dickhoffbook,mahaux91,dussan14}
\begin{equation}
S^{p(c)}_{l_\Lambda j_\Lambda}(k_\Lambda, k'_\Lambda, \omega)=
\lim_{\eta\to 0^+}
\frac{i}{2\pi}
\left(
g^{\Lambda}_{l_\Lambda j_\Lambda}(k_\Lambda,k'_\Lambda, \omega+i\eta)
-
g^{\Lambda}_{l_\Lambda j_\Lambda}(k_\Lambda,k'_\Lambda, \omega-i\eta) 
\right) \ . 
\label{eq:sflc}
\end{equation}
Note that using the Lehmann representation of the propagator it is easy to show that Eq.\ (\ref{eq:sflc}) can be simply rewritten as
\begin{equation}
S^{p(c)}_{l_\Lambda j_\Lambda}(k_\Lambda, k'_\Lambda, \omega)=
-\frac{1}{\pi}\mbox{Im}\, g^{\Lambda}_{l_\Lambda j_\Lambda}(k_\Lambda,k'_\Lambda, \omega) \ .
\label{eq:sflc2}
\end{equation}

Due to the presence of the delta function in Eq.\ (\ref{eq:slp2}), however, it is numerically more convenient to obtain the continuum contribution of $\Lambda$ spectral function in coordinate space. Using Eqs.\ (\ref{eq:slp2}) and (\ref{eq:sflc2}) in the Fourier--Bessel transform
\begin{equation}
S^{p(c)}_{l_\Lambda j_\Lambda}(r_\Lambda, r'_\Lambda, \omega)=
\frac{2}{\pi}\int_{0}^{\infty}dk_\Lambda k_\Lambda^2
\int_{0}^{\infty} dk'_\Lambda k'^2_\Lambda
j_{l_\Lambda}(k_\Lambda r_\Lambda)
S^{p(c)}_{l_\Lambda j_\Lambda}(k_\Lambda, k'_\Lambda, \omega)
j_{l_\Lambda}(k'_\Lambda r'_\Lambda) 
\label{eq:fbtsf}
\end{equation}
we obtain 
\begin{eqnarray}
S^{p(c)}_{l_\Lambda j_\Lambda}(r_\Lambda, r'_\Lambda, \omega)&=&
\frac{2}{\pi}\frac{m_\Lambda k_0}{\hbar^2}j_{l_\Lambda}(k_0 r_\Lambda)j_{l_\Lambda}(k_0 r'_\Lambda) \nonumber \\
&+&
2\left(\frac{m_\Lambda k_0}{\hbar^2}\right)^2j_{l_\Lambda}(k_0 r_\Lambda)j_{l_\Lambda}(k_0 r'_\Lambda)
\mbox{Im}\, \Sigma^{red}_{l_\Lambda j_\Lambda}(k_0,k_0, \omega) \nonumber \\
&+&
\frac{2}{\pi}\frac{m_\Lambda k_0}{\hbar^2}j_{l_\Lambda}(k_0 r_\Lambda)
{\cal P}\int_{0}^{\infty} dk'_\Lambda k'^2_\Lambda 
\frac{j_{l_\Lambda}(k'_\Lambda r'_\Lambda)\mbox{Re}\,\Sigma^{red}_{l_\Lambda j_\Lambda}(k_0,k'_\Lambda, \omega)}{\omega-\frac{\hbar^2k'^2_\Lambda}{2m_\Lambda}} \nonumber \\
&+&
\frac{2}{\pi}\frac{m_\Lambda k_0}{\hbar^2}j_{l_\Lambda}(k_0 r'_\Lambda)
{\cal P}\int_{0}^{\infty} dk_\Lambda k^2_\Lambda 
\frac{j_{l_\Lambda}(k_\Lambda r_\Lambda)\mbox{Re}\,\Sigma^{red}_{l_\Lambda j_\Lambda}(k_\Lambda,k_0, \omega)}{\omega-\frac{\hbar^2k^2_\Lambda}{2m_\Lambda}} \nonumber \\
&-&
\frac{2}{\pi^2}
{\cal P}\int_{0}^{\infty} dk_\Lambda k^2_\Lambda
\frac{j_{l_\Lambda}(k_\Lambda r_\Lambda)}{\omega-\frac{\hbar^2k^2_\Lambda}{2m_\Lambda}} 
{\cal P}\int_{0}^{\infty} dk'_\Lambda k'^2_\Lambda 
\frac{j_{l_\Lambda}(k'_\Lambda r'_\Lambda)\mbox{Im}\,\Sigma^{red}_{l_\Lambda j_\Lambda}(k_\Lambda,k'_\Lambda, \omega)}{\omega-\frac{\hbar^2k'^2_\Lambda}{2m_\Lambda}} \nonumber \\ 
\label{eq:fbtsf}
\end{eqnarray}
where $k_0=\sqrt{2m_\Lambda \omega}/\hbar$ and ${\cal P}$ denotes a principal value integral.

The spectral strenght in the continuum of a $\Lambda$ in the $l_\Lambda  j_\Lambda$ partial wave
can be then obtained from the following double folding of the spectral function  \cite{dussan14}
\begin{equation}
S^{p(c)}_{l_\Lambda  j_\Lambda}(\omega)=
\int_{0}^{\infty}dr_\Lambda r_\Lambda^2 \int_{0}^{\infty} dr'_\Lambda r'^2_\Lambda
\Psi_{l_\Lambda j_\Lambda}(r_\Lambda)
S^{p(c)}_{l_\Lambda j_\Lambda}(r_\Lambda, r'_\Lambda, \omega)
\Psi_{l_\Lambda j_\Lambda}(r'_\Lambda) \ .
\label{eq:ssflc}
\end{equation}

Then, the total spectral strength is finally given by the sum of the discrete and continuum contributions
\begin{equation}
S^{p}_{l_\Lambda  j_\Lambda}(\omega)=S^{p(d)}_{l_\Lambda  j_\Lambda}(\omega)+S^{p(c)}_{l_\Lambda  j_\Lambda}(\omega) \ .
\label{eq:tss}
\end{equation}



\begin{figure}[t]
\begin{center}
\includegraphics[width=15.0cm]{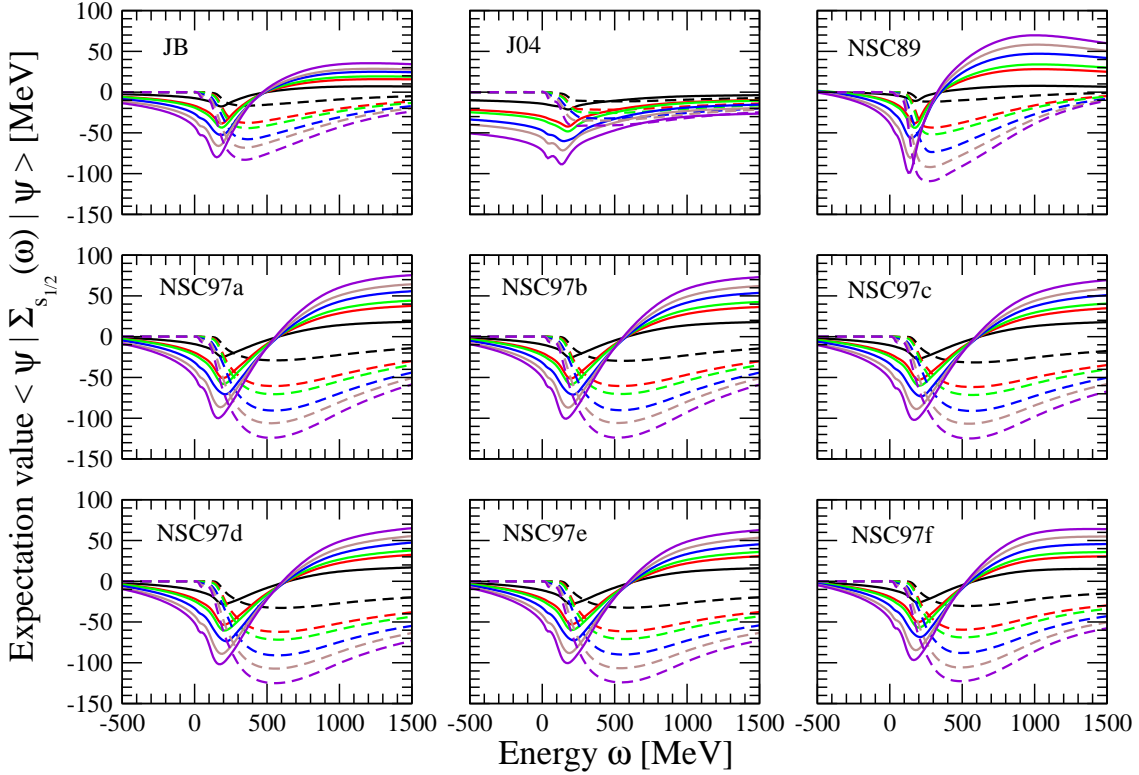}
\caption{(color online) Energy dependence of the real $\langle \Psi | {\cal V}_{l_\Lambda j_\Lambda}|\Psi \rangle$ (solid lines) and imaginary $\langle \Psi | {\cal W}_{l_\Lambda j_\Lambda}|\Psi \rangle$ (dashed lines) part of the expectation value of the self-energy of a $\Lambda$ with $l_\Lambda=0$ and $j_\Lambda=1/2$ 
for $^5_{\Lambda}$He (black), $^{13}_{\,\,\,\Lambda}$C (red), $^{17}_{\,\,\,\Lambda}$O (green), $^{41}_{\,\,\,\Lambda}$Ca (blue), $^{91}_{\,\,\,\Lambda}$Zr (brown) and $^{209}_{\,\,\,\,\,\Lambda}$Pb (violet) 
obtained with the different YN interactions. The energy is measured with respect to the $\Lambda$ rest mass.}
\label{fig:selfener2}
\end{center}
\end{figure}

\section{Results}
\label{sec:res}

In this section we present our results for the self-energy, single-particle bound states and the spectral function of a $\Lambda$ in $^5_{\Lambda}$He, $^{13}_{\,\,\,\Lambda}$C, $^{17}_{\,\,\,\Lambda}$O, $^{41}_{\,\,\,\Lambda}$Ca, $^{91}_{\,\,\,\Lambda}$Zr and $^{209}_{\,\,\,\,\,\Lambda}$Pb. The results have been obtained using some of the J\"{u}lich and Nijmegen YN interactions; namely, the models J\"{u}lich B (JB) \cite{juelich} and J\"{u}lich 04 (J04) \cite{juelich2}, and the Nijmegen soft-core models NSC89 \cite{nijmegen} and NSC97a-f \cite{nijmegen2,nijmegen3}. Results for the disoccupation of $\Lambda$ bound and scattering states are also presented and discussed at the end of the section. 

\begin{figure}[t]
\begin{center}
\includegraphics[width=15.0cm]{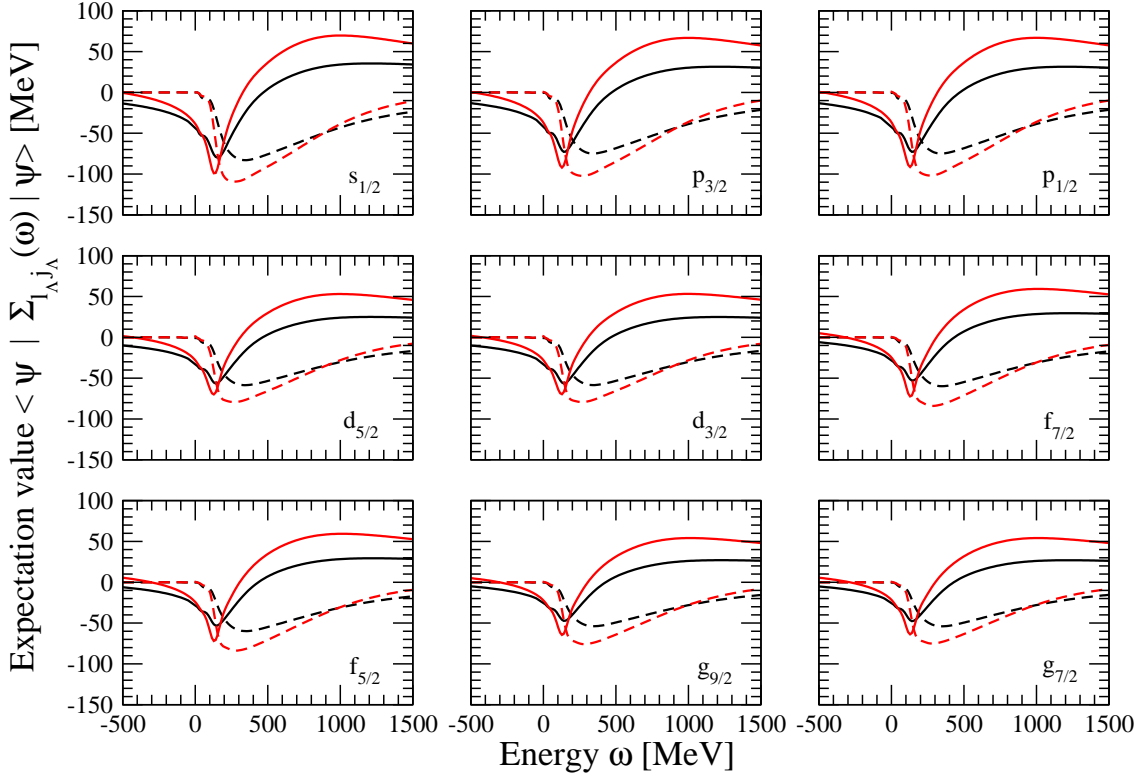}
\caption{(color online) Energy dependence of the real $\langle \Psi | {\cal V}_{l_\Lambda j_\Lambda}|\Psi \rangle$ (solid lines) and imaginary $\langle \Psi | {\cal W}_{l_\Lambda j_\Lambda}|\Psi \rangle$ (dashed lines) part of the expectation value of the self-energy of a $\Lambda$ in the $s, p, d, f$ and $g$ partial waves 
for $^{209}_{\,\,\,\,\,\Lambda}$Pb obtained with the JB (black lines) and NSC89 (red lines) models. The energy is measured with respect to the $\Lambda$ rest mass.}
\label{fig:selfener3}
\end{center}
\end{figure}

\subsection{$\Lambda$ self-energy}

We start by showing in Fig.\ \ref{fig:selfener2} the energy dependence of the real 
$\langle \Psi | {\cal V}_{l_\Lambda j_\Lambda}|\Psi \rangle$ and imaginary  
$\langle \Psi | {\cal W}_{l_\Lambda j_\Lambda}|\Psi \rangle$ 
part of the expectation value (see Eqs.\ (\ref{eq:lse}) and (\ref{eq:ev})) of the self-energy of a $\Lambda$ with $l_\Lambda=0$ and $j_\Lambda=1/2$, for the six hypernuclei considered.  Results are presented for all the YN interactions mentioned above. As it is seen in the figure, in general, 
$\langle \Psi | {\cal W}_{l_\Lambda j_\Lambda}|\Psi \rangle$ is, in absolute value, larger in the Nijmegen models than in the J\"{u}lich ones. Consequently, the dispersion relation of Eq.\ (\ref{eq:diagc2}) leads in the case of the Nijmegen models to a stronger energy dependence of $\langle \Psi | {\cal V}_{l_\Lambda j_\Lambda}|\Psi \rangle$. We observe that $\langle \Psi | {\cal W}_{l_\Lambda j_\Lambda}|\Psi \rangle$ is only different from zero for $\omega > 0$ and that it is always negative, even for energies larger than those shown in the figure. Moreover, due to phase space restrictions, $\langle \Psi | {\cal W}_{l_\Lambda j_\Lambda}|\Psi \rangle$ behaves almost quadratically for energies close to $\omega=0$ \cite{luttinger61,sartor77}
\begin{equation}
\langle \Psi | {\cal W}_{l_\Lambda j_\Lambda}| \Psi \rangle \sim c\, \omega^2, \,\,\,\,\,\,\,\,\, \omega \longrightarrow 0^+ \ .
\label{eq:asymp}
\end{equation}

The energy dependence of $\langle \Psi | {\cal V}_{l_\Lambda j_\Lambda}|\Psi \rangle$ can be understood from the dispersion relation (\ref{eq:diagc2}). Since $\langle \Psi | {\cal W}_{l_\Lambda j_\Lambda}|\Psi \rangle$ is different from zero only for $\omega > 0$ and it is negative, this dispersion relation implies that $\langle \Psi | {\cal V}_{l_\Lambda j_\Lambda}|\Psi \rangle$ will be attractive for $\omega < 0$. This attraction increases up to a given positive value of $\omega$ and then it decreases, eventually turning into repulsion 
at large energies. Note that, up to $\omega \sim 500-600$ MeV, $\langle \Psi | {\cal V}_{l_\Lambda j_\Lambda}|\Psi \rangle$ is more attractive for the heavier hypernuclei becoming more repulsive than that of the lighter ones for higher energies.

All these features can be also observed in Fig.\ \ref{fig:selfener3} where, for completeness, we show the energy dependence of $\langle \Psi | {\cal V}_{l_\Lambda j_\Lambda}|\Psi \rangle$ and $\langle \Psi | {\cal W}_{l_\Lambda j_\Lambda}|\Psi \rangle$ of a $\Lambda$ in the $s, p, d, f$ and $g$ partial waves 
for $^{209}_{\,\,\,\,\,\Lambda}$Pb. The results are shown for the JB and  NSC89 models. Note that the dependence of both $\langle \Psi | {\cal V}_{l_\Lambda j_\Lambda}|\Psi \rangle$ and $\langle \Psi | {\cal W}_{l_\Lambda j_\Lambda}|\Psi \rangle$ on the orbital ($l_\Lambda$) and total ($j_\Lambda$) angular momentum of the $\Lambda$ is rather weak.


\begin{table*}[t]
\begin{center}
\scriptsize
\begin{tabular}{c|c|ccccccccc|c}
\hline
\hline
Nuclei & $l_\Lambda j_\Lambda$ & JB & J04 & NSC89 & NSC97a & NSC97b & NSC97c & NSC97d & NSC97e & NSC97f & Exp. \\
\hline
$^5_{\Lambda}$He &  &  &  & &  &  &  &   &  & &  ($^5_{\Lambda}$He) \\ 
                              & $s_{1/2}$ & $-2.28$ & $-5.89$ & $-0.58$& $-3.16$ & $-3.38$ & $-3.94$ &  $-4.24$ & $-4.20$ & $-3.59$ & $-3.12$  \\ 
\hline
$^{13}_{\,\,\,\Lambda}$C &  &  &  & &  &  &  &   &  & &  ($^{13}_{\,\,\,\Lambda}$C) \\
& $s_{1/2}$ & $-9.48$& $-18.94$ & $-5.69$ & $-11.46$& $-11.79$ & $-12.76$ & $-13.08$ & $-12.82$ & $-11.37$ & $-11.69$\\
& $p_{3/2}$ & $-$       & $-3.66$  &  $-$       &  $-0.24$  & $-0.32$  &  $-0.63$  & $-0.68$   &  $-0.54$   & $-0.01$  & $-0.7$ (p)\\
& $p_{1/2}$ & $-$       & $-4.07$  &  $-$       &  $-0.12$  & $-0.14$  &  $-0.37$  & $-0.35$   &  $-0.19$   & $-$  & \\
\hline
$^{17}_{\,\,\,\Lambda}$O &  &  &  & &  &  &  &   &  & &  ($^{16}_{\,\,\,\Lambda}$O) \\
& $s_{1/2}$ & $-11.83$& $-23.40$ & $-7.39$ & $-14.31$& $-14.65$ & $-15.70$ & $-15.99$ & $-15.68$ & $-14.02$ & $-12.5$\\
& $p_{3/2}$ & $-0.87$  & $-8.16$  &  $-$       &  $-2.57$  & $-2.72$  &  $-3.24$  & $-3.33$   &  $-3.10$   & $-2.17$  & $-2.5$ (p)\\
& $p_{1/2}$ & $-1.06$  & $-8.03$  &  $-$       &  $-2.16$  & $-2.22$  &  $-2.61$  & $-2.57$   &  $-2.30$   & $-1.41$  & \\
\hline
$^{41}_{\,\,\,\Lambda}$Ca &  &  &  & &  &  &  &   &  & &  ($^{40}_{\,\,\,\Lambda}$Ca) \\
& $s_{1/2}$ & $-19.60$ & $-36.16$ & $-15.04$  & $-23.09$ & $-23.42$  &$-24.60$ &  $-24.74$ & $-24.20$ & $-21.96$ &  $-20.0$\\ 
& $p_{3/2}$ & $-9.64$ & $-23.81$ & $-6.92$   & $-12.37$ & $-12.57$   &$-13.40$ & $-13.35$  & $-12.84$&$-11.09$ & $-12.0$ (p)\\
& $p_{1/2}$ & $-9.92$ & $-23.78$ & $-6.29$   & $-12.10$   &$-12.23$    &$-12.95$ & $-12.78$  & $-12.22$ &  $-10.45$&\\
& $ d_{5/2}$ & $-0.70$ & $-11.72$ & $-$            & $-2.80$   &$-2.93$     &$-3.47$    & $-3.38$    & $-3.00$ & $-1.83$ & $-1.0$ (d) \\
& $d_{3/2}$ & $-1.01$ & $-11.65$ & $-$            & $-2.43$   &$-2.46$     &$-2.85$    &$-2.61$     &  $-2.18$& $-1.04$ &  \\
\hline
$^{91}_{\,\,\,\Lambda}$Zr &  &  &  & &  &  &  &   &  & &  ($^{89}_{\,\,\,\Lambda}$Y) \\
& $s_{1/2}$ &$-25.80$ &$-46.30$ &$-22.77$ &$-31.38$ &$-31.73$ &$-33.05$  &$-33.06$ &$-32.33$ &$-29.56$ &  $-23.0$\\ 
& $p_{3/2}$ &$-18.19$ &$-37.73$ &$-17.08$ &$-23.92$ &$-24.20$ &$-25.28$  &$-25.22$ &$-24.58$ &$-22.25$ & $-16.0$ (p)\\
& $p_{1/2}$ &$-18.30$ &$-38.01$ &$-16.68$ &$-23.82$ &$-24.06$ &$-25.07$  &$-24.92$ &$-24.23$ &$-21.88$ & \\
& $d_{5/2}$ &$-11.16$ &$-28.35$ &$-9.05$ &$-14.41$  &$-14.58$ &$-15.36$ &$-15.09$ &$-14.42$ &$-12.41$  & $-9.0$ (d)\\
& $d_{3/2}$ &$-11.17$ &$-28.44$ &$-8.49$ &$-14.30$  &$-14.40$ &$-15.12$ &$-14.77$ &$-14.06$ &$-11.99$ &  \\
& $f_{7/2}$ &$-3.05$    &$-18.45$ &$-1.56$ &$-5.46$   &$-5.52$ &$-6.03$        &$-5.59$ &$-4.93$ &$-3.27$  & $-2.0$ (f)\\
& $f_{5/2}$ &$-2.99$    &$-18.76$ &$-1.00$ &$-5.28$   &$-5.26$ &$-5.69$        &$-5.20$ &$-4.52$ &$-2.86$  & \\
\hline
$^{209}_{\,\,\,\,\,\Lambda}$Pb &  &  &  & &  &  &  &   &  & &  ($^{208}_{\,\,\,\,\,\Lambda}$Pb) \\
& $s_{1/2}$  &$-31.36$ &$-59.95$ &$-29.52$ &$-38.85$ &$-39.23$ &$-40.63$ &$-40.44$ &$-39.50$ &$-39.30$ & $-27.0$\\ 
& $p_{3/2}$ &$-27.13$  &$-55.21$ &$-26.01$ &$-33.49$ &$-33.91$ &$-35.13$ &$-34.80$ &$-33.86$ &$-31.03$  & $-22.0$ (p) \\
& $p_{1/2}$ &$-27.18$  &$-55.40$ &$-25.72$ &$-33.38$ &$-33.78$ &$-34.94$ &$-34.54$ &$-33.56$ &$-30.72$ & \\
& $d_{5/2}$ &$-21.70$  &$-45.08$ &$-17.85$ &$-23.23$ &$-23.54$ &$-24.38$ &$-23.79$ &$-22.858$ &$-20.60$  & $-17.0$ (d) \\
& $d_{3/2}$ &$-21.77$  &$-45.07$ &$-17.65$ &$-23.17$ &$-23.45$ &$-24.27$ &$-23.68$ &$-22.75$ &$-20.51$  & \\
& $f_{7/2}$ &$-13.00$   &$-37.15$ &$-9.67$ &$-15.38$ &$-15.43$ &$-16.04$ &$-15.05$ &$-13.81$ &$-10.98$  & $-12.0$ (f)\\
& $f_{5/2}$ &$-13.13$   &$-37.16$ &$-9.31$ &$-15.35$ &$-15.33$ &$-15.90$ &$-14.87$ &$-13.61$ &$-10.76$  & \\
& $g_{9/2}$ &$-8.14$    &$-29.91$ &$-5.27$ &$-10.07$ &$-10.14$ &$-10.68$ &$-9.80$ &$-8.71$ &$-6.28$  & $-7.0$ (g) \\
& $g_{7/2}$ &$-8.26$   &$-30.16$ &$-4.80$ &$-10.01$ &$-10.00$ &$-10.46$ &$-9.49$ &$-8.37$ & $-5.91$  & \\
\hline
\hline
\end{tabular}
\end{center}
\caption{Energy of $\Lambda$ single-particle bound states of several hypernuclei from $^5_{\Lambda}$He to $^{209}_{\,\,\,\,\,\Lambda}$Pb
for the different YN interactions considered. Available experimental, data taken from Refs.\ \cite{bando90,pile91,hasegawa96}, is shown for the closest measured hypernuclei.
Units are given in MeV.}
\label{tab:energ}
\end{table*}


\begin{figure}[t]
\begin{center}
\includegraphics[width=15.0cm]{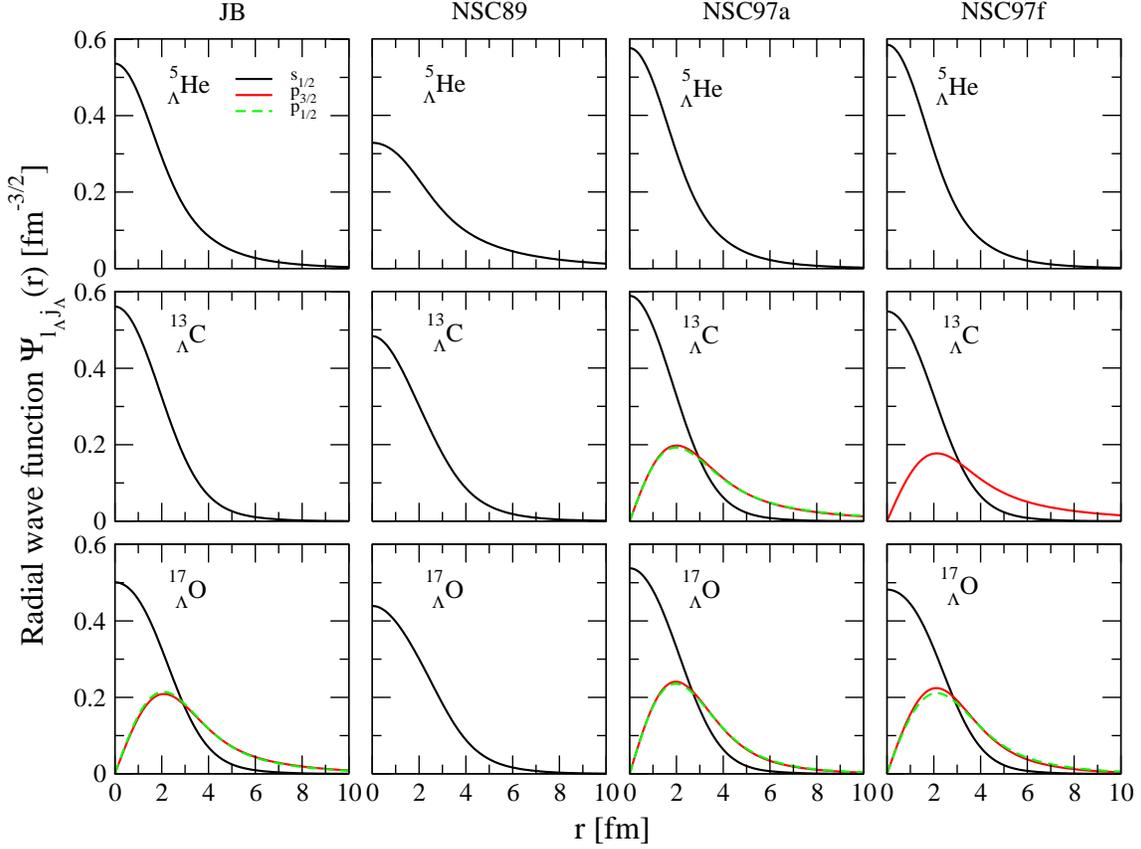}
\caption{(color online) Radial wave function for the $s-$ and $p-$wave states of a $\Lambda$ hyperon in $^5_{\Lambda}$He (upper panels), $^{13}_{\,\,\,\Lambda}$C (middle panels) and $^{17}_{\,\,\,\Lambda}$O
(lower panels) predicted by the JB, NSC89, NSC97a and NSC97f models.}
\label{fig:wf1}
\end{center}
\end{figure}

\begin{figure}[t]
\begin{center}
\includegraphics[width=15.0cm]{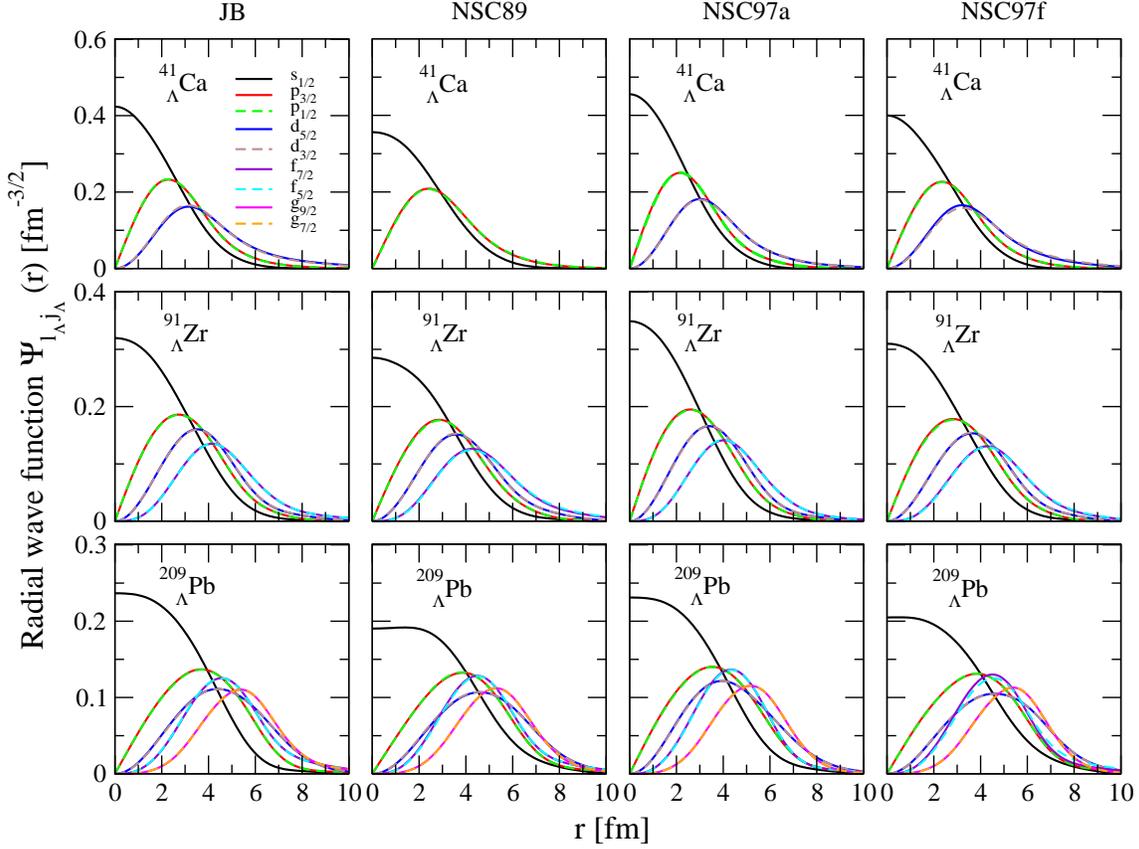}
\caption{(color online) Radial wave function for the $s-, p-, d-, f-$ and $g-$wave states of a $\Lambda$ hyperon in 
$^{41}_{\,\,\,\Lambda}$Ca  (upper panels), $^{91}_{\,\,\,\Lambda}$Zr (middle panels) and $^{209}_{\,\,\,\,\,\Lambda}$Pb (lower panels)
predicted by the JB, NSC89, NSC97a and NSC97f models.}
\label{fig:wf2}
\end{center}
\end{figure}

\subsection{$\Lambda$ single-particle bound states}

The energy of $\Lambda$ single-particle bound states in $^5_{\Lambda}$He, $^{13}_{\,\,\,\Lambda}$C, $^{17}_{\,\,\,\Lambda}$O, $^{41}_{\,\,\,\Lambda}$Ca, $^{91}_{\,\,\,\Lambda}$Zr and $^{209}_{\,\,\,\,\,\Lambda}$Pb
for the different YN interactions considered in this work are shown in Tab.\ \ref{tab:energ}. Note that 
we have considered only hypernuclei that are described as a closed shell nuclear core plus a $\Lambda$ sitting in a single-particle state. Since experimental data for those hypernuclei do not always exist, we show for comparison the closest representative hypernuclei for which experimental information is available. Experimental data have been taken from Refs.\ \cite{bando90,pile91,hasegawa96}. We note, however, that the differences between the calculated and the experimental values should not be associated only to this fact but mainly to the approximations made in the calculation or to the uncertainties of the YN interactions employed. 

In general the agreement with experimental data is qualitatively good for most of the models except for the J04 one, which for all hypernuclei predicts an unrealistic overbinding of the $\Lambda$ in all the 
single-particle states. For this reason, in the following we will not present further results for this model. Note that the results for $^{91}_{\,\,\,\Lambda}$Zr and $^{209}_{\,\,\,\,\,\Lambda}$Pb appear clearly overbound also for the other models, specially for the NSC97a-f ones. This overbinding is mainly due to the fact that 
the NSC97a-f models are too much attractive, predicting a $\Lambda$ single-particle potential in symmetric nuclear matter of about $- 40$ MeV,  in comparison with the value of around $-30$ MeV extrapolated from hypernuclear data \cite{skhf1}. We note also that the distortion of the plane wave associated with the nucleon in the intermediate state of Fig.\ \ref{fig:selfener}c, which is needed to guarantee its orthogonalization with the nucleon hole state (see {\it e.g.,} Ref.\ \cite{borromeo92} for a detail description), has been done in an approximate way. In addition, the orthogonalization procedure has been optimized only for the case of $^{17}_{\,\,\,\Lambda}$O. Furthermore, in the calculation of the $\Lambda$ self-energy only short-range correlations have been taken into account by means of the ladder diagrams where the intermediate states, as it is shown in Sec.\ \ref{sec:sf}, are treated in an approximate way (see Eq.\ (\ref{eq:g2nd})). The effect of long-range correlations or the coupling to collective excitations \cite{barbieri04}, which eventually could improve the agreement with the experimental data, have been ignored in the present calculation. Note finally that the splitting of the $p-, d-, f-$ and $g-$wave states is of the order of few tenths of MeV in all cases due to the small spin-orbit strength of the YN interaction.

For completeness we plot in Figs.\ \ref{fig:wf1} and \ref{fig:wf2} the radial wave function of the $\Lambda$ in the different bound states of the six hypernuclei considered. The results are shown only for the JB, NSC89, NSC97a and NSC97f YN interactions. Results for the NSC97b-e models are not shown because their difference with respect to those of the NSC97a and NSC97f is smaller than $\sim 10\%$. Therefore, from now on we will consider the NSC97a and NSC97f models together with the NSC89 one as representative of the Nijmegen Soft Core YN interaction. 

The probability of finding the $\Lambda$ at the center of the hypernucleus is given by $|\Psi_{s_{1/2}}(r=0)|^2$. As it can bee seen in Figs.\ \ref{fig:wf1} and \ref{fig:wf2}, going from light to heavy hypernuclei the wave function of the $s_{1/2}$ state becomes more and more spread due to the larger extension of the nuclear density over which the $\Lambda$ hyperon wants to be distributed and, therefore, this probability  decreases. Only the light hypernucleus $^5_{\Lambda}$He falls out of this pattern due to the fact that the energy of the bound state $s_{1/2}$ is very small in this case, therefore, resulting in a very extended wave function, specially for the NSC89 model. Note now from Tab.\ \ref{tab:energ} that, in fact, the NSC89 model predicts always the larger single-particle energies 
which result in the larger radii for all the bound $\Lambda$ states and, therefore, in more extended wave functions than those predicted by the other models which are more localized. 
This is illustrated in particular for the $s_{1/2}$ state in Tab.\ \ref{tab:rad}, where we show its root-mean-square radius for the different hypernuclei. Finally, we would like to conclude this discussion by noticing that, due to the scale of the figures, the small spin-orbit splitting of the $p-, d-, f-$ and $g-$wave states cannot be resolved in the corresponding wave functions.


\begin{table*}[t]
\begin{center}
\small
\begin{tabular}{c|cccc}
\hline
\hline
Nuclei  & JB  & NSC89 & NSC97a  & NSC97f  \\
\hline
$^5_{\Lambda}$He & $3.08$  & $4.83$  & $2.79$ & $2.70$  \\
\hline
$^{13}_{\,\,\,\Lambda}$C  & $2.43$ & $2.79$ & $2.34$ & $2.40$ \\
\hline
$^{17}_{\,\,\,\Lambda}$O & $2.47$ & $2.80$ & $2.39$ & $2.47$   \\
\hline
$^{41}_{\,\,\,\Lambda}$Ca & $2.75$ & $2.99$ & $2.66$ & $2.83$   \\
\hline
$^{91}_{\,\,\,\Lambda}$Zr  & $3.15$ & $3.31$ & $3.01$ & $3.20$  \\
\hline
$^{209}_{\,\,\,\,\,\Lambda}$Pb & $3.55$ & $3.96$ & $3.65$ & $3.96$   \\
\hline
\hline
\end{tabular}
\end{center}
\caption{Root-mean-square radius  of the $\Lambda$ $s_{1/2}$ bound state in
several hypernuclei from $^5_{\Lambda}$He to $^{209}_{\,\,\,\,\,\Lambda}$Pb for the  JB, NSC89, NSC97a and NSC97f models. Units are given in fm.}
\label{tab:rad}
\end{table*}


\begin{figure}[t]
\begin{center}
\includegraphics[width=10.0cm]{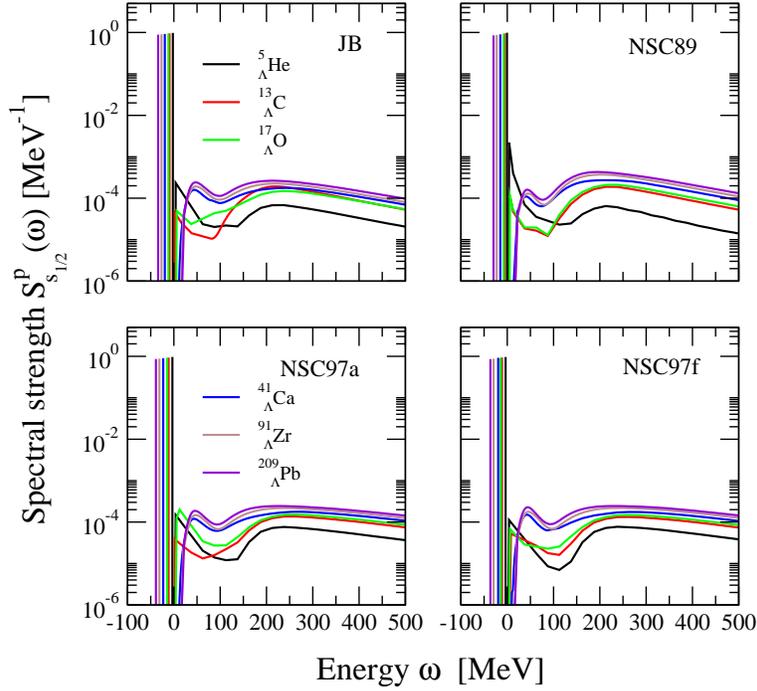}
\caption{(color online) Total spectral strength of a
$\Lambda$ with $l_\Lambda=0$ and $j_\Lambda=1/2$
for $^5_{\Lambda}$He (black), $^{13}_{\,\,\,\Lambda}$C (red), $^{17}_{\,\,\,\Lambda}$O (green), $^{41}_{\,\,\,\Lambda}$Ca (blue), $^{91}_{\,\,\,\Lambda}$Zr (brown) and $^{209}_{\,\,\,\,\,\Lambda}$Pb (violet) 
predicted by the JB, NSC89, NSc97a and NSC97f models. For each hypernuclei and model,
the discrete contribution is shown by a weighted delta function located at the corresponding energy of the $s_{1/2}$ bound state. The contribution from the continuum is spread over all positive energies. The energy is measured with respect to the $\Lambda$ rest mass.}
\label{fig:sf1}
\end{center}
\end{figure}

\subsection{$\Lambda$ spectral function}

In Fig.\ \ref{fig:sf1} we show the energy dependence of the total spectral strength (Eq.\ (\ref{eq:tss})) of a $\Lambda$ with $l_\Lambda=0$ and $j_\Lambda=1/2$ for all the hypernuclei considered. The results have been obtained with the JB, NSC89, NSC97a and NSC97f YN interactions. The discrete contribution, as it was said in Sec.\ \ref{sec:sf}, is a delta function located at the energy of the single-particle $s_{1/2}$ bound state of the corresponding hypernuclei (see Tab.\ \ref{tab:energ}) whose strength is given by the factor $Z_{l_\Lambda j_\Lambda}$ (see Eq.\ (\ref{eq:ssfld2})). On the other hand, the strength of the continuum contribution (Eq.\ (\ref{eq:ssflc})) is spread over all positive energies. As it is seen in the figure, this contribution shows some structure for $\omega \lesssim 100$ MeV, reflection of the behaviour of the self-energy in this energy region, and it decreases monotonically for $\omega \gtrsim 200$ MeV. We note that the strength of the discrete contribution decreases when one moves from light to heavier hypernuclei. This can be easily understood by noticing (see Eq.\ (\ref{eq:specf}) and Fig.\ \ref{fig:selfener2}) that the derivative of $\langle \Psi | \Sigma_{s_{1/2}}(\omega)|\Psi \rangle$ with respect to $\omega$ at the bound energy $\varepsilon_{l_\Lambda j_\Lambda}$ is real, negative and larger in absolute value for the heavier hypernuclei. All this reflects the fact that the $\Lambda$-nucleon correlations become more and more important when the density of the nuclear core increases, and it shows the fact that the smaller the value of $Z_{l_\Lambda j_\Lambda}$ is, the more important are the correlations of the system.  We note, however, that the value of $Z_{l_\Lambda j_\Lambda}$, shown in Tab.\  \ref{tab:sfac} for the different $\Lambda$ single-particle bound states, is in general relatively large for all hypernuclei. This is in agreement with the fact, already mentioned in the introduction, that the $\Lambda$ keeps its identity iside the nucleus and that is less correlated than the nucleons. This was already observed in nuclear matter by Robertson and Dickhoff in Refs.\ \cite{robertson04, robertson04b}.   

\begin{figure}[t]
\begin{center}
\includegraphics[width=15.0cm]{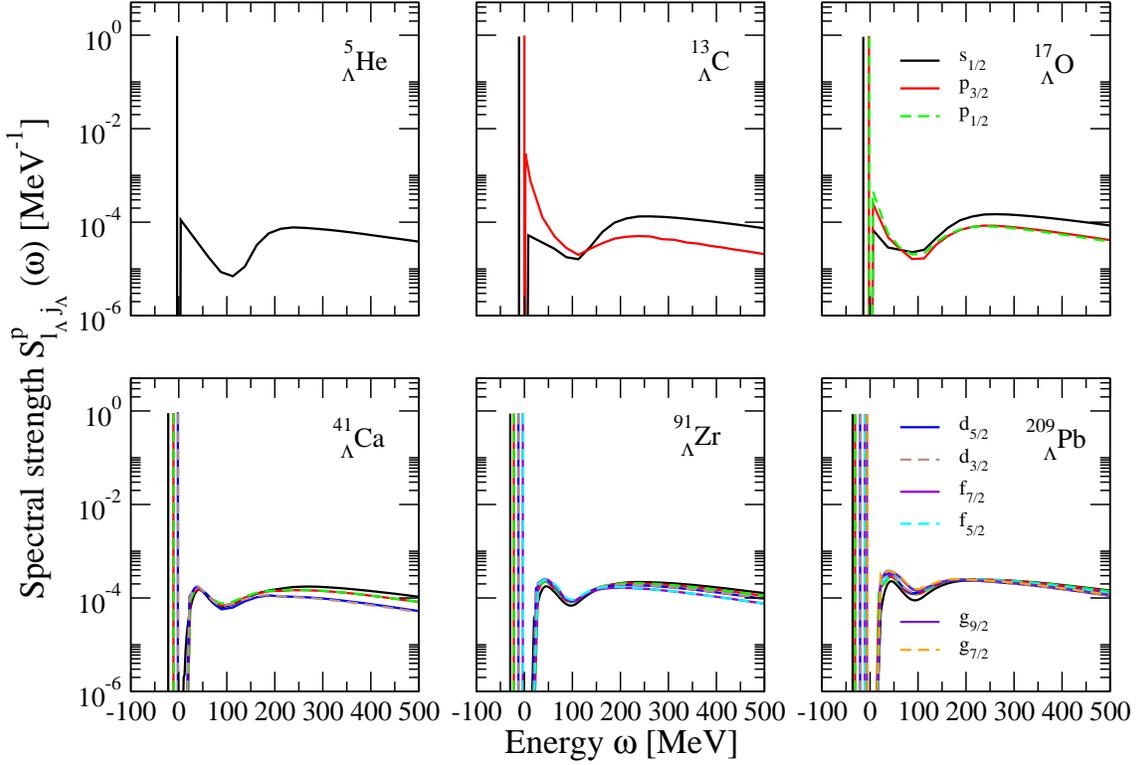}
\caption{(color online)
Total spectral strength of a
$\Lambda$ in the
$s, p, d, f$ and $g$ partial waves for several hypernuclei predicted by the NSC97f model. As in Fig.\ \ref{fig:sf1} 
the discrete contributions are shown by weighted delta functions located at the corresponding energies of the different  bound states. The contribution from the continuum is spread over 
all positive energies. The energy is measured with respect to the $\Lambda$ rest mass.}
\label{fig:sf2}
\end{center}
\end{figure}

We finish this part of the section by showing in Fig.\ \ref{fig:sf2} the total spectral strength of a $\Lambda$ in the $s, p, d, f$ and $g$ partial waves for several hypernuclei predicted by the NSC97f model. The figure presents general features very similar to the ones just described and, therefore, there is no need to discuss them again. Note simply that the strength of the discrete contribution increases when increasing the partial wave $l_\Lambda j_\Lambda$ (see Tab.\  \ref{tab:sfac}), indicating that the $\Lambda$-nucleon correlations become less important for the higher partial waves. Note also that, for the heavier hypernuclei, the strength of the contrinuum contribution becomes very similar for all partial waves at high energies. This is again easily understood from the energy and the $l_\Lambda j_\Lambda$ dependence of $\langle \Psi | \Sigma_{l_\Lambda j_\Lambda}(\omega)|\Psi \rangle$ (see Fig.\ \ref{fig:selfener3}). 


\begin{table*}[t]
\begin{center}
\scriptsize
\begin{tabular}{c|c|cccc}
\hline
\hline
Nuclei & $l_\Lambda j_\Lambda$ & JB & NSC89 & NSC97a & NSC97f  \\
\hline
$^5_{\Lambda}$He &  &  &  &  &    \\ 
                              & $s_{1/2}$ & $0.976$ & $0.983$ & $0.965$ & $0.964$   \\ 
\hline

$^{13}_{\,\,\,\Lambda}$C  &  &  &  &  &  \\ 
& $s_{1/2}$ & $0.950$ & $0.940$ & $0.933$ & $0.933$   \\ 
& $p_{3/2}$ & $-$ & $-$ & $0.975$ & $0.979$   \\
& $p_{1/2}$ & $-$ & $-$ & $0.976$ & $$   \\
\hline
$^{17}_{\,\,\,\Lambda}$O &  &  &  &  &  \\  
& $s_{1/2}$ & $0.942$ & $0.930$ & $0.923$ & $0.924$   \\
& $p_{3/2}$ & $0.973$ & $-$ & $0.956$ & $0.959$   \\
& $p_{1/2}$ & $0.971$ & $-$ & $0.957$ & $0.961$   \\
\hline
$^{41}_{\,\,\,\Lambda}$Ca &  &  &  &  &  \\  
& $s_{1/2}$ & $0.920$ & $0.896$ & $0.898$ & $0.898$   \\
& $p_{3/2}$ & $0.930$ & $0.915$ & $0.911$ & $0.914$   \\
& $p_{1/2}$ & $0.929$ & $0.914$ & $0.910$ & $0.912$   \\
& $ d_{5/2}$ & $0.952$ & $-$ & $0.932$ & $0.938$   \\
& $d_{3/2}$ & $0.949$ & $-$ & $0.931$ & $0.939$   \\
\hline
$^{91}_{\,\,\,\Lambda}$Zr &  &  &  &  &  \\ 
& $s_{1/2}$ & $0.904$ & $0.870$ & $0.879$ & $0.876$   \\ 
& $p_{3/2}$ & $0.906$ & $0.875$ & $0.884$ & $0.883$   \\
& $p_{1/2}$ & $0.907$ & $0.876$ & $0.885$ & $0.883$   \\
& $d_{5/2}$ & $0.910$ & $0.886$ & $0.891$ & $0.893$   \\
& $d_{3/2}$ & $0.911$ & $0.886$ & $0.891$ & $0.891$   \\
& $f_{7/2}$ & $0.919$ & $0.903$ & $0.903$ & $0.906$   \\
& $f_{5/2}$ & $0.920$ & $0.905$ & $0.902$ & $0.907$   \\
\hline
$^{209}_{\,\,\,\,\,\Lambda}$Pb &  &  &  &  &  \\ 
& $s_{1/2}$  & $0.884$ & $0.846$ & $0.857$ & $0.856$   \\ 
& $p_{3/2}$ & $0.885$ & $0.847$ & $0.858$ & $0.857$   \\
& $p_{1/2}$ & $0.885$ & $0.847$ & $0.858$ & $0.857$   \\
& $d_{5/2}$ & $0.896$ & $0.858$ & $0.870$ & $0.869$   \\
& $d_{3/2}$ & $0.896$ & $0.857$ & $0.869$ & $0.867$   \\
& $f_{7/2}$ & $0.891$ & $0.852$ & $0.863$ & $0.857$   \\
& $f_{5/2}$ & $0.891$ & $0.851$ & $0.863$ & $0.855$   \\
& $g_{9/2}$ & $0.892$ & $0.855$ & $0.869$ & $0.862$   \\
& $g_{7/2}$ & $0.892$ & $0.854$ & $0.868$ & $0.860$   \\
\hline
\hline
\end{tabular}
\end{center}
\caption{Factor $Z_{l_\Lambda j_\Lambda }$ for the $\Lambda$ single-particle bound states in several hypernuclei from $^5_{\Lambda}$He to $^{209}_{\,\,\,\,\,\Lambda}$Pb
predicted by the JB, NSC89, NSC97a and NSC97f models.}
\label{tab:sfac}
\end{table*}


\subsection{Disoccupation number}

\begin{figure}[t]
\begin{center}
\includegraphics[width=15.0cm]{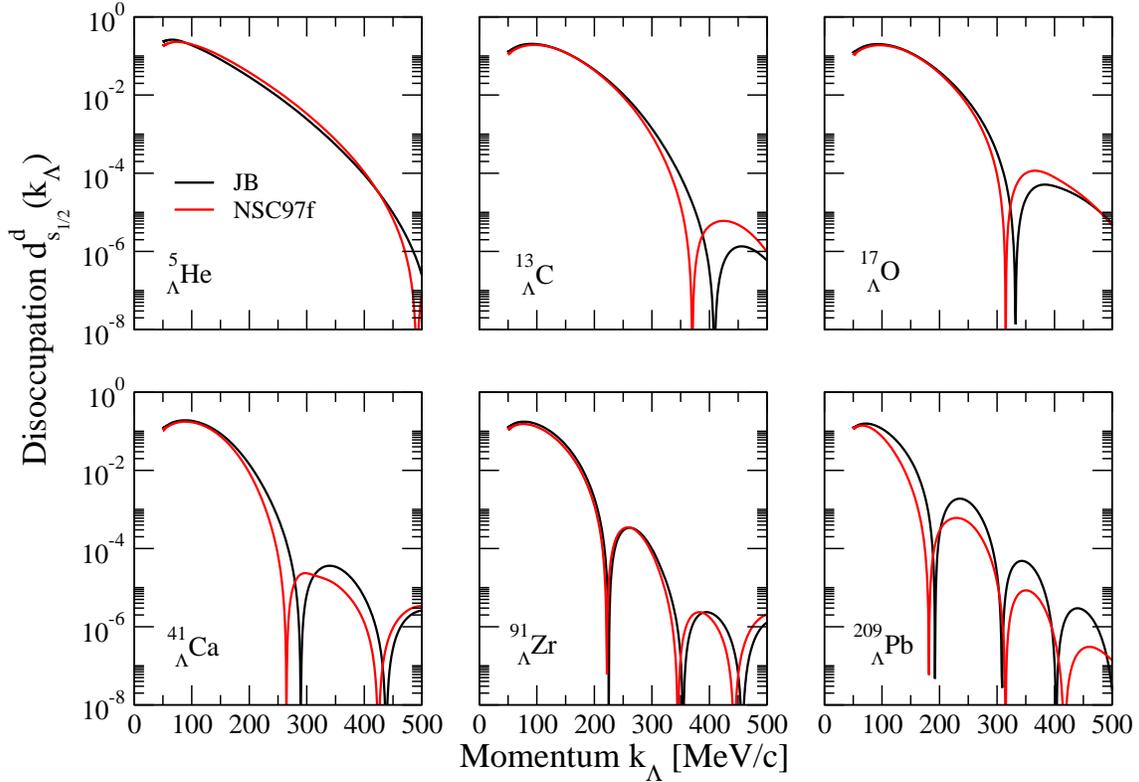}
\caption{(color online) Discrete contribution to the disoccupation of the $s_{1/2}$ state in different hypernuclei from $^5_{\Lambda}$He to $^{209}_{\,\,\,\,\,\Lambda}$Pb obtained with the JB (black lines) and NSC97f (red lines) models.}
\label{fig:dis1}
\end{center}
\end{figure}

\begin{figure}[t]
\begin{center}
\includegraphics[width=15.0cm]{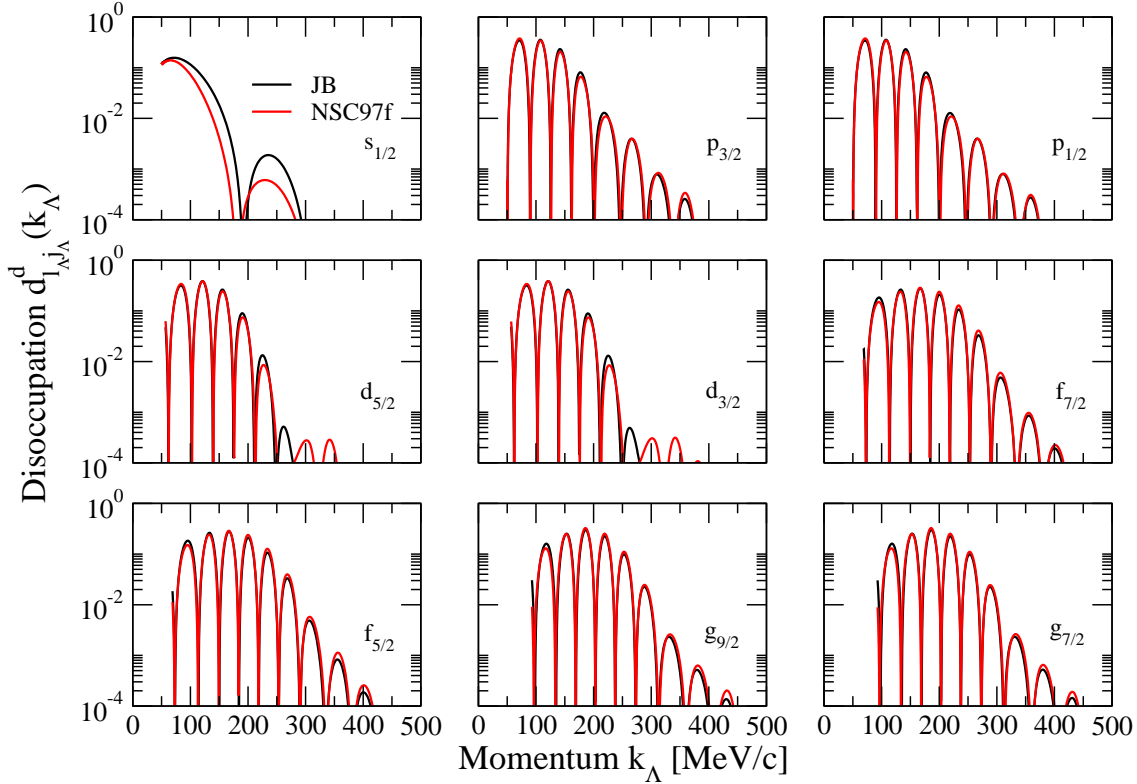}
\caption{(color online) Discrete contribution to the disoccupation of the $s-, p-, d-, f-$  and $g-$wave states in $^{209}_{\,\,\,\,\,\Lambda}$Pb obtained with the JB (black lines) and NSC97f (red lines) models.}
\label{fig:dis2}
\end{center}
\end{figure}

We show in Figs.\ \ref{fig:dis1} and \ref{fig:dis2}, respectively, the discrete contribution to the disocupation of the $s_{1/2}$ state in the different hypernuclei, and that of the $s-, p-, d-, f-$  and $g-$wave ones in $^{209}_{\,\,\,\,\,\Lambda}$Pb, predicted by the JB and the NSC97f models, obtained by integrating Eq.\  (\ref{eq:sfld}) over the energy
\begin{equation}
d^{d}_{l_\Lambda  j_\Lambda}(k_\Lambda)=\int_{\mu_\Lambda}^{\infty}d\omega S^{p(d)}_{l_\Lambda j_\Lambda}(k_\Lambda, \omega)
= Z_{l_\Lambda j_\Lambda}
|\langle k_\Lambda l_\Lambda j_\Lambda m_{j_\Lambda}| \Psi \rangle|^2 \ ,
\label{eq:dkd}
\end{equation}
where 
\begin{equation}
\mu_\Lambda=E(^{A+1}_{\,\,\,\,\,\,\,\,\,\Lambda} Z) - E(^AZ) \ ,
\end{equation}
is the $\Lambda$ chemical potential being $E(^{A+1}_{\,\,\,\,\,\,\,\,\,\Lambda} Z)$ and $E(^AZ)$ the ground state energies of the hypernucleus $^{A+1}_{\,\,\,\,\,\,\,\,\,\Lambda} Z$ and the nucleus $^AZ$, respectively. The quantity $d^{d}_{l_\Lambda  j_\Lambda}(k_\Lambda)$ gives the probability of adding a $\Lambda$ of momentum $k_\Lambda$ in the single-particle bound state $l_\Lambda j_\Lambda$ of the hypernucleus. Intuitively, one expects that if the momentum $k_\Lambda$ is large, then the $\Lambda$ can easily escape and, therefore, the probability of binding it in the nucleus should be quite small. This feature is clearly shown in both figures, where we can observe that, in fact, $d^{d}_{l_\Lambda  j_\Lambda}(k_\Lambda)$ decreases when increasing $k_\Lambda$ and becomes almost negligible for large values of the momentum of the $\Lambda$. This indicates that in hypernuclear production reactions the $\Lambda$ hyperon is formed mostly in a quasi-free state. We note that the oscillation behaviour of $d^{d}_{l_\Lambda  j_\Lambda}(k_\Lambda)$ is simply due to the zeros of the projection coefficient $\langle k_\Lambda l_\Lambda j_\Lambda | \Psi \rangle$, that in the logarithmic scale of the plot appear as singularities. 

We would like to finish this section by mentioning that the total spectral strength of the $\Lambda$ hyperon fulfills the following sum rule
\begin{equation}
\int_{\mu_\Lambda}^{\infty}d\omega S^{p}_{l_\Lambda j_\Lambda}(\omega) = 1
\label{eq:sr}
\end{equation}
which simply express that the total disoccupation number is $1$ or, in other words, that it is always possible to add a $\Lambda$ with quantum numbers $l_\Lambda j_\Lambda$ either in a single-particle bound state or in a scattering state of a given ordinary nucleus.  Eq.\ (\ref{eq:sr}) can be splitted as a sum of a discrete and a continuum contribution
\begin{equation}
\int_{\mu_\Lambda}^{\infty}d\omega S^{p}_{l_\Lambda j_\Lambda}(\omega)=
\int_{\mu_\Lambda}^{\infty}d\omega S^{p(d)}_{l_\Lambda j_\Lambda}(\omega)+
\int_{\mu_\Lambda}^{\infty}d\omega S^{p(c)}_{l_\Lambda j_\Lambda}(\omega)=1  \ ,
\label{eq:totp}
\end{equation}
where the discrete one gives just the factor $Z_{l_\Lambda j_\Lambda}$. In Tab.\ \ref{tab:prob} we show the result of both contributions as well as their sum for the different states of all the hypernuclei considered predicted by the NSC97f YN interaction. We note that the integration of the continuum contribution have been done up to $500$ MeV which, in part, can explain why the sum of both contributions is smaller than $1$.

\begin{table*}[t]
\begin{center}
\scriptsize
\begin{tabular}{c|lccccccccc}
\hline
\hline
Nuclei &  & $s_{1/2}$ & $p_{3/2}$ & $p_{1/2}$ & $d_{5/2}$ & $d_{3/2}$ & $f_{7/2}$ & $f_{5/2}$ & $g_{9/2}$ & $g_{7/2}$  \\
\hline
$^5_{\Lambda}$He &  & $$ & $$ & $$ & $$ & $$ & $$ & $$ & $$ & $$  \\
                               & Discrete & $0.964$ & $-$ & $-$ & $-$ & $-$ & $-$ & $-$ & $-$ & $-$  \\
                               & Continuum & $0.023$ & $-$ & $-$ & $-$ & $-$ & $-$ & $-$ & $-$ & $-$  \\
                               & Total & $0.987$ & $-$ & $-$ & $-$ & $-$ & $-$ & $-$ & $-$ & $-$  \\
\hline
$^{13}_{\,\,\,\Lambda}$C &  & $$ & $$ & $$ & $$ & $$ & $$ & $$ & $$ & $$  \\
                               & Discrete & $0.933$ & $0.979$ & $-$ & $-$ & $-$ & $-$ & $-$ & $-$ & $-$  \\
                               & Continuum & $0.040$ & $0.017$ & $-$ & $-$ & $-$ & $-$ & $-$ & $-$ & $-$  \\
                               & Total & $0.973$ & $0.996$ & $-$ & $-$ & $-$ & $-$ & $-$ & $-$ & $-$  \\
\hline
$^{17}_{\,\,\,\Lambda}$O &  & $$ & $$ & $$ & $$ & $$ & $$ & $$ & $$ & $$  \\
                               & Discrete & $0.924$ & $0.959$ & $0.961$ & $-$ & $-$ & $-$ & $-$ & $-$ & $-$  \\
                               & Continuum & $0.053$ & $0.037$ & $0.036$ & $-$ & $-$ & $-$ & $-$ & $-$ & $-$  \\
                               & Total & $0.977$ & $0.996$ & $0.997$ & $-$ & $-$ & $-$ & $-$ & $-$ & $-$  \\
\hline
$^{41}_{\,\,\,\Lambda}$Ca &  & $$ & $$ & $$ & $$ & $$ & $$ & $$ & $$ & $$  \\
                               & Discrete & $0.898$ & $0.914$ & $0.912$ & $0.938$ & $0.939$ & $-$ & $-$ & $-$ & $-$  \\
                               & Continuum & $0.071$ & $0.063$ & $0.064$ & $0.048$ & $0.047$ & $-$ & $-$ & $-$ & $-$  \\
                               & Total & $0.969$ & $0.977$ & $0.976$ & $0.986$ & $0.986$ & $-$ & $-$ & $-$ & $-$  \\
\hline
$^{91}_{\,\,\,\Lambda}$Zr &  & $$ & $$ & $$ & $$ & $$ & $$ & $$ & $$ & $$  \\
                               & Discrete & $0.876$ & $0.883$ & $0.883$ & $0.893$ & $0.891$ & $0.906$ & $0.907$ & $-$ & $-$  \\
                               & Continuum & $0.120$ & $0.113$ & $0.113$ & $0.103$ & $0.105$ & $0.089$ & $0.090$ & $-$ & $-$  \\
                               & Total & $0.996$ & $0.996$ & $0.996$ & $0.996$ & $0.996$ & $0.995$ & $0.997$ & $-$ & $-$  \\
\hline
$^{209}_{\,\,\,\,\,\Lambda}$Pb &  & $$ & $$ & $$ & $$ & $$ & $$ & $$ & $$ & $$  \\
                               & Discrete     & $0.856$ & $0.857$ & $0.857$ & $0.869$ & $0.867$ & $0.857$ & $0.855$ & $0.862$ & $0.860$  \\
                               & Continuum & $0.138$ & $0.142$ & $0.142$ & $0.129$ & $0.130$ & $0.140$ & $0.141$ & $0.137$ & $0.139$  \\
                               & Total          & $0.994$ & $0.999$ & $0.999$ & $0.998$ & $0.997$ & $0.997$ & $0.996$ & $0.999$ & $0.999$  \\
\hline
\hline
\end{tabular}
\end{center}
\caption{Discrete, continuum contribution and total disoccupation number of the different states for all the hypernuclei considered predicted by the NSC97f YN interaction.}
\label{tab:prob}
\end{table*}



\section{Summary and conclusions}
\label{sec:sumcon}

In this work we have determined the single-particle spectral function of the $\Lambda$ hyperon in several hypernuclei. To such end, we have obtained first the corresponding 
$\Lambda$ self-energy using a perturbative many-body approach with some of the YN interactions of the J\"{u}lich \cite{juelich,juelich2} and the Nijmegen \cite{nijmegen,nijmegen2,nijmegen3} groups.
The calculation started with the construction of a nuclear matter YN $G$-matrix that was used to build a finite nucleus one through a perturbative expansion that is truncated at second order. This finite nuclei YN $G$-matrix was then employed to calculate the $\Lambda$ self-energy. From it we obtained finally the $\Lambda$ spectral function and the binding energies, wave functions and disoccupation numbers of different single-particle states for various hypernuclei from $^5_{\Lambda}$He to $^{209}_{\,\,\,\,\,\Lambda}$Pb. 

We showed that the Nijmegen models predict an imaginary part of the $\Lambda$ self-energy larger than that obtained with the J\"{u}lich ones, which results in a stronger energy dependence 
of its real part. The dependence of the $\Lambda$ self-energy on the orbital ($l_\Lambda$) and total ($j_\Lambda$) angular momentum  of the $\Lambda$ was found to be quite small.

Using the real part of the $\Lambda$ self-energy as an effective hyperon-nucleus potential in the Schr\"{o}dinger equation, we obtained the $\Lambda$ single-particle bound states in the several hypernuclei considered. Our results compared rather well with the experimental data for all the models except for the J04, which predicted an unrealistic overbinding of all the $\Lambda$ single-particle orbits. The small spin-orbit splitting of the $p-, d-, f-$ and $g-$wave states was confirmed.

The discrete and the continuum contributions of the total spectral strength of the $\Lambda$ hyperon were then obtained from the $\Lambda$ self-energy. Their appearance was found to be qualitatively similar to that of the nucleons. We showed that the discrete contribution is a weigthed delta function located at the energy of the corresponding single-$\Lambda$ bound state, and that the strength of the continuum one is spread over all positive energies. The factor $Z_{l_\Lambda j_\Lambda}$, that measures the importance of the correlations, was also calculated for the $s_{1/2}$ state. Our results showed that the value of $Z_{l_\Lambda j_\Lambda}$ was relatively large, indicating that the $\Lambda$ is less correlated than the nucleons, in agreement with the idea that it mantains its identity inside the nucleus, and with the results of a previous study of the $\Lambda$ correlations in infinite nuclear matter \cite{robertson04,robertson04b}.

Finally, by integrating the $\Lambda$ spectral function over the energy, we obtained the disoccupation numbers of different single-particle states for the various hypernuclei. Our results showed that the discrete contribution to the disoccupation number decreases when increasing the momentum of the $\Lambda$, indicating that in hypernuclear production reactions the $\Lambda$ hyperon is formed mostly in a quasi-free state. To finish, we would like to say that scattering reactions, such as the high precision $(e,e'K^+)$ ones carried out at JLAB and MAMI-C, can provide valuable information on the discoccupation of $\Lambda$ single-particle bound states needed to have a  more complete description of the properties of the $\Lambda$ hyperon in nuclear systems.


\section*{Acknowledgements}

The author wants to express his deep gratitude to Artur Polls for all the useful and stimulating conversations they had during the development of this work. This work is supported by the 
NewCompstar, COST Action MP1304.




\end{document}